\documentclass[prb,twocolumn,aps,showpacs,letter]{revtex4}

\usepackage{graphicx}% complex graphics
\usepackage{bm}% bold math
\usepackage{amssymb} %math symbols
\usepackage{amsmath}
\usepackage{subfigure}
\usepackage{widetext}
\usepackage{bigints}

\begin{document}
\newcommand{\s}{\scriptscriptstyle}
\newcommand{\be}{\begin{equation}}
\newcommand{\ee}{\end{equation}}

%%%%%%%%%%%%%%%%%%%%%%%%%%%%%%%%%%%%%%%%%%%%%%%%%%%%%%%%%%%%%%%%%%%%%%%%%%%%%%%%%%%%%%%%%%%%%%%%%%%%%%%%%%%%%%%%%%%%%%%%%%%%%%%%%%%%%%%%%%%%%%%%%%%%%%%%%%%%%%%%%%%%%%%%%%%%%%%%%%%%%%%%%%%%%%% %%%%%%%%%%%%%%%%%%%%%%%%%%%%%%%%%%%%%%%%%%%%%%%%%%%%%%%%%%%%%%%%%%%%%%%%%%%%%%%%%%%%%%%%%%%%%%%%%%%%%%%%%%%%%%%%%%%%%%%%%%%%%%%%%%%%%%%%%%%%%%%%%%%%%%%%%%%%%%%%%%%%%%%%%%%%%%%%%%%%%%%%%%%%%%% %%%%%%%%%%%%%%%%%%%%%%%%%%%%%%%%%%%%%%%%%%%%%%%%%%%%%%%%%%%%%%%%%%%%%%%%%%%%%%%%%%%%%%%%%%%%%%%%%%%%%%%%%%%%%%%%%%%%%%%%%%%%%%%%%%%%%%%%%%%%%%%%%%%%%%%%%%%%%%%%%%%%%%%%%%%%%%%%%%%%%%%%%%%%%%% %%%%%%%%%%%%%%%%%%%%%%%%%%%%%%%%%%%%%%%%%%%%%%%%%%%%%%%%%%%%%%%%%%%%%%%%%%%%%%%%%%%%%%%%%%%%%%%%%%%%%%%%%%%%%%%%%%%%%%%%%%%%%%%%%%%%%%%%%%%%%%%%%%%%%%%%%%%%%%%%%%%%%%%%%%%%%%%%%%%%%%%%%%%%%%%
\title{Analytical description of spin--Rabi oscillation controlled electronic transitions rates between weakly coupled pairs of paramagnetic states with S=1/2}

\author{R. Glenn, W. J. Baker, C. Boehme, and M. E. Raikh }
%\date{}
\affiliation{Department of Physics and Astronomy, University of Utah, Salt Lake City, UT 84112}
\begin{abstract}
    We report on an analytical description of spin--dependent electronic transition rates which are controlled by a radiation induced spin-Rabi oscillation of weakly spin--exchange and spin--dipolar coupled paramagnetic states ($S=\frac{1}{2}$). The oscillation components (the Fourier content) of the net transition rates within spin--pair ensembles are derived for randomly distributed spin resonances with account of a possible correlation between the two distributions that correspond to the two individual pair partners. The results presented here show that when electrically or optically detected Rabi spectroscopy is conducted under an increasing driving field $B_{\s 1}$, the Rabi spectrum evolves from a single resonance peak at $s=\Omega_{{\scriptscriptstyle R}}$, where $\Omega_{{\scriptscriptstyle R}}=\gamma B_{\s 1}$ is the Rabi frequency ($\gamma$ is the gyromagnetic ratio), to {\em three} peaks at $s= \Omega_{{\scriptscriptstyle R}}$, $s=2\Omega_{{\scriptscriptstyle R}}$, and at low $s\ll \Omega_{{\scriptscriptstyle R}}$. The crossover between the two regimes takes place when $\Omega_{{\scriptscriptstyle R}}$ exceeds the expectation value $\delta_{{\scriptscriptstyle 0}}$ of the difference of the Zeeman energies within the pairs, which corresponds to the broadening of the magnetic resonance lines in the presence of disorder caused by hyperfine field or distributions of Land\'e $g$-factors. We capture this crossover by analytically calculating the shapes of all three peaks at arbitrary relation between $\Omega_{{\scriptscriptstyle R}}$ and $\delta_{{\scriptscriptstyle 0}}$. When the peaks are well-developed their widths are ${\s \Delta} s \sim \delta_{{\scriptscriptstyle 0}}^2/\Omega_{{\scriptscriptstyle R}}$.
\end{abstract}
%76.30.-v  Electron paramagnetic resonance and relaxation}
%42.50.Md 	Optical transient phenomena: quantum beats, photon echo, free-induction decay, dephasings and revivals, optical nutation, and self-induced transparency
%71.35.Gg 	Exciton-mediated interactions

\maketitle

%%%%%%%%%%%%%%%%%%%%%%%%%%%%%%%%%%%%%%%%%%%%%%%%%%%%%%%%%%%%%%%%%%%%%%%%%%%%%%%%%%%%%%%%%%%%%%%%%%%%%%%%%%%%%%%%%%%%%%%%%%%%%%%%%%%%%%%%%%%%%%%%%%%%%%%%%%%%%%%%%%%%%%%%%%%%%%%%%%%%%%%%%%%%%%% %%%%%%%%%%%%%%%%%%%%%%%%%%%%%%%%%%%%%%%%%%%%%%%%%%%%%%%%%%%%%%%%%%%%%%%%%%%%%%%%%%%%%%%%%%%%%%%%%%%%%%%%%%%%%%%%%%%%%%%%%%%%%%%%%%%%%%%%%%%%%%%%%%%%%%%%%%%%%%%%%%%%%%%%%%%%%%%%%%%%%%%%%%%%%%% %%%%%%%%%%%%%%%%%%%%%%%%%%%%%%%%%%%%%%%%%%%%%%%%%%%%%%%%%%%%%%%%%%%%%%%%%%%%%%%%%%%%%%%%%%%%%%%%%%%%%%%%%%%%%%%%%%%%%%%%%%%%%%%%%%%%%%%%%%%%%%%%%%%%%%%%%%%%%%%%%%%%%%%%%%%%%%%%%%%%%%%%%%%%%%% %%%%%%%%%%%%%%%%%%%%%%%%%%%%%%%%%%%%%%%%%%%%%%%%%%%%%%%%%%%%%%%%%%%%%%%%%%%%%%%%%%%%%%%%%%%%%%%%%%%%%%%%%%%%%%%%%%%%%%%%%%%%%%%%%%%%%%%%%%%%%%%%%%%%%%%%%%%%%%%%%%%%%%%%%%%%%%%%%%%%%%%%%%%%%%%
\section{Introduction}
Over the past decade, pulsed electrically detected magnetic resonance (pEDMR) spectroscopy has been used increasingly for the investigation of the physical nature of spin--dependent electronic transitions which influence conductivity such as excess charge carrier recombination or transport transitions through localized paramagnetic states as seen in amorphous inorganic~\cite{PhysRevB.79.195205, lee_192104}, crystalline~\cite{Boehme2003b, Stegner2006, PhysRevLett.106.187601, PhysRevLett.100.177602} as well as organic~\cite{BoehmeC60, BoehmeNature, schaefer2008electrical, BoehmeDifferentiation, wrong} semiconductor materials. In most of these experimental studies, pEDMR experiments are conducted within a pulse--probe scheme~\cite{BoehmeMain, CPHC_CPHC201000186} where the current through the host materials of the spin systems of interest is measured while short and intensive magnetic resonant pulses are imposed. The pulsed magnetic resonant radiation prepares coherent spin non-eigenstates from initial, well defined eigenstates before the pulse. As the changed spin--states will also cause changes of the spin--dependent conductivity, one can gain information about the prepared coherent spin--state by integration of the electric current transient after the pulse which eventually, on long time scales (compared to the length of the coherent excitation), will return to its pre--pulse steady state. The probed charge (obtained from the integrated current transient) depends on the coherent spin state after the pulse which in turn depends on the pulse parameters (length, frequency, intensity)~\cite{BoehmeMain}. Therefore, a measurement of the charge as a function of the applied pulse length can reveal the propagation of a spin state in presence of the resonant radiation pulse. Thus, current detectable observations of spin Rabi oscillations are possible.

%%%%%%%%%%%%%%%%%%%%%%%%%%%%%%%%%%%%%%%%%%%%%%%%%%%%%%%%%%%%%%%%%%%%%%%%%%%%%%%%%%%%%%%%%%%%%%%%%%%%%%%%%%%%%%%%%%%%%%%%%%%%%%%%%%%%%%%%%%%%%%%%%%%%%%%%%%%%%%%%%%%%%%%%%%%%%%%%%%%%%%%%%%%%%%%
\begin{figure}
\includegraphics[width=77mm, angle=0,clip]{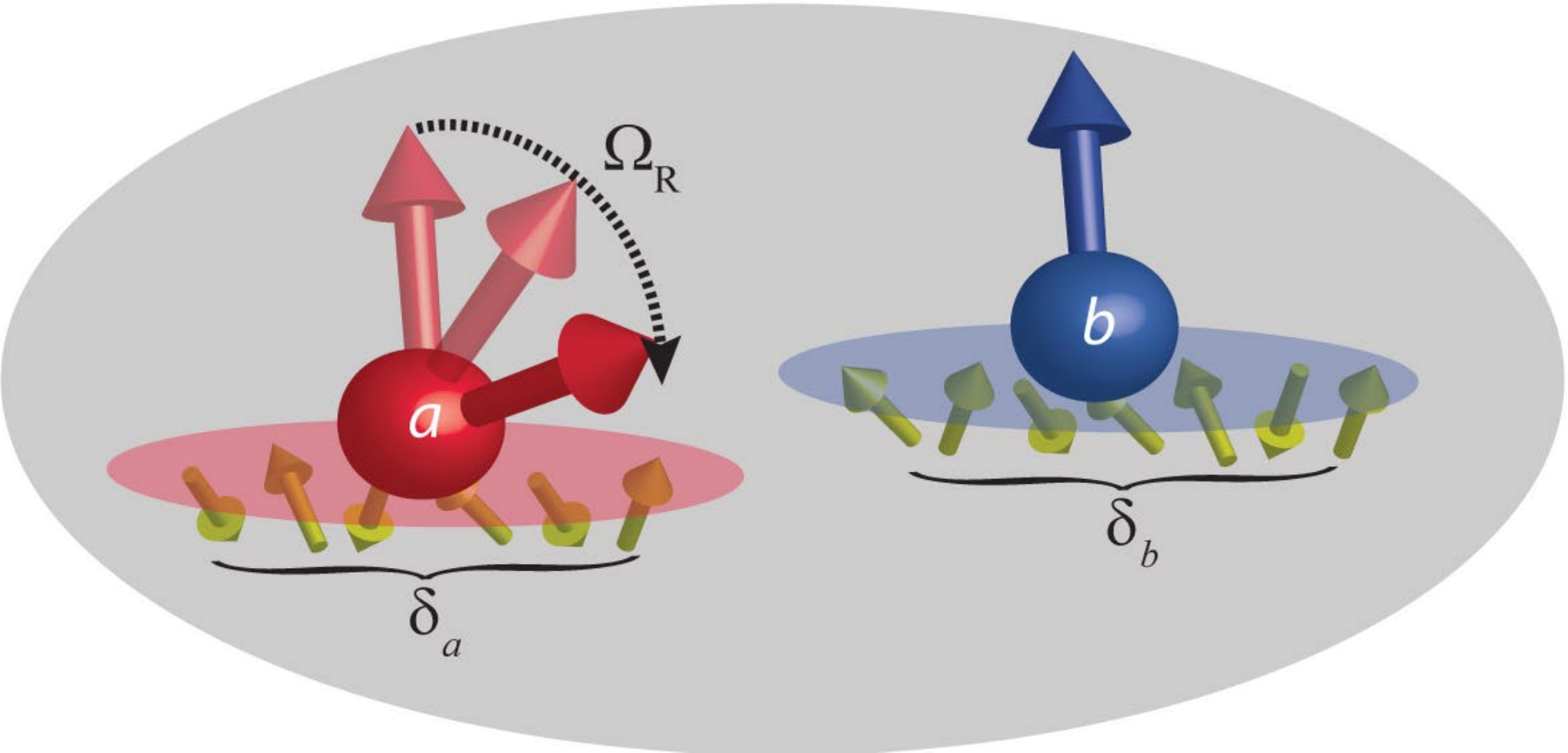}
\vspace{-0.4cm}
\caption{Schematic illustration of the Rabi oscillations with frequency $\Omega_{\s R}$ in spin-$\frac{1}{2}$
pair. Components of the pair, $a$ and $b$ have different environments
causing random shifts, $\delta_{\s a}$ and $\delta_{\s b}$, from the resonant
frequency, $\omega_{\s 0}$.  Relevant for PEDMR are the initial and final spin configurations  $|\!\uparrow \uparrow\rangle$ and $|\!\downarrow\downarrow\rangle$ only.}
\label{pair}
\end{figure}
%%%%%%%%%%%%%%%%%%%%%%%%%%%%%%%%%%%%%%%%%%%%%%%%%%%%%%%%%%%%%%%%%%%%%%%%%%%%%%%%%%%%%%%%%%%%%%%%%%%%%%%%%%%%%%%%%%%%%%%%%%%%%%%%%%%%%%%%%%%%%%%%%%%%%%%%%%%%%%%%%%%%%%%%%%%%%%%%%%%%%%%%%%%%%%%

Most EDMR detectable spin--dependent electronic transitions reported in the literature are due to Pauli--blockade effects which occur for transitions between two paramagnetic states with $S=\frac{1}{2}$. These systems, illustrated in Fig.~\ref{pair}, usually require weak spin--orbit coupling as found in materials with low atomic order numbers (this means silicon and carbon materials) as well as sufficiently weak spin--spin coupling (which means exchange and dipolar interaction) within the formed pairs. In order to allow one of the electrons within this pair of paramagnetic states to undergo a transition into the other paramagnetic state (thereby forming a singlet spin manifold due to the Pauli exclusion principle), the spin pair state $|\psi\rangle$ before the transition requires non--negligible singlet content ($\langle\psi|S\rangle\neq0$) for the transition to have a non--negligible probability. The special nature of such intermediate pair controlled spin--dependent transition rates was first recognized by Kaplan, Solomon and Mott~\cite{Mott} who explained the magnitude of continuous wave EDMR experiments at the time. With the advent of pEDMR about a decade ago, this model also became most significant for the understanding of many of the EDMR detected coherent spin motion experiments.

When pEDMR is applied to the intermediate pair processes described by Kaplan, Solomon and Mott, the observable applied to the spin ensemble is permutation symmetry (the singlet operator), it is not the magnetic polarization of the spin ensemble as it is the case for conventional magnetic resonance spectroscopies which are based on the detection of radiation. Some implications of this observable change have been discussed theoretically in previous studies on intermediate pairs for cases of no intrapair interactions~\cite{BoehmeMain,Boehme1}, cases of weak but non--negligible exchange interaction~\cite{Boehme2} (weak here means that the exchange interaction is much smaller than the spin--Zeeman splitting of the pair partners but not necessarily weaker than the difference of the pair partners Larmor frequencies), and for cases where disorder within the spin ensemble~\cite{Boehme3} is significant. These numerical studies have shown that an electrically detectable spin--Rabi oscillation can contain various harmonic components which can essentially form a "fingerprint" for the spin--Hamiltonian of the observed pairs. Thus, conducting a Fourier analysis of an observed spin--Rabi signal (one could call this Rabi spectroscopy) can give microscopic information about the nature of charge carrier states or of paramagnetic defects.

Most of the previously published pEDMR studies have been conducted as Rabi spectroscopy experiments~\cite{Stegner2006, lee_192104, BoehmeC60, BoehmeNature, wrong, BoehmeDifferentiation}. For most of these experimental data, a correct interpretation would be impossible without the information provided by the existing theoretical studies~\cite{BoehmeMain,Boehme1,Boehme2,Boehme3}. Nevertheless, these studies can only provide limited support for an experimental analysis since numerical simulations can only provide answers about the behavior of a simulated system for a fixed set of parameters, but they do not reveal analytical expressions that can be fit or directly compared to experimental data, and most importantly, they oftentimes do not enhance the qualitative understanding of a simulated system. It has been the goal of this study to overcome this problem by finding a closed analytical form for the description of spin--Rabi oscillation controlled spin--dependent transition rates within spin pairs with $S=\frac{1}{2}$. The expressions derived in the following reveal the dependence of all harmonic components found with electrically, and similarly, with intermediate spin--pair controlled optically detected transition rates on the parameters of both, the observed physical system as well as the performed Rabi--oscillation experiment.

The parameters of a performed Rabi--oscillation experiment are the spin $S=\frac{1}{2}$ Rabi oscillation frequency $\Omega_{{\scriptscriptstyle R}}=\gamma B_1$ which depends on the driving field strength $B_1$ as well as the gyromagnetic ratio $\gamma$. The parameters characterizing the observed spin pair are given by the spin--orbit controlled respective $g$-factors for each pair partner as well as local, and hyperfine fields which in general are different for the two pair partners as well. For our purposes, all these parameters can be taken into account by the difference $\delta$ of the pair partners Larmor frequencies. Taking this into account, we anticipate from the previously reported numerical simulations~\cite{BoehmeMain, Boehme1} that in the weak--driving regime where $\Omega_{{\scriptscriptstyle R}}\ll\delta$, the Fourier transformation ${\mathcal {\bf F}} (s)$ of a Rabi oscillation transient exhibits only one peak at $s=\Omega_{\s R}$, while in the strong-driving regime $\Omega_{{\scriptscriptstyle R}}\gg\delta$) there is one peak at $s=2\Omega_{\s R}$ and no peak at $s=\Omega_{\s R}$. The crossover between these two situations occurs around $\Omega_{{\scriptscriptstyle R}}\approx\delta$.

This crossover with increasing $\Omega_{\s R}$ was demonstrated experimentally for spin--dependent polaron pair recombination processes in different organic semiconductor materials~\cite{BoehmeDifferentiation, wrong}. Currently, it is rather difficult (if not impossible) to extract quantitative information from such experimentally measured Rabi spectra ${\mathcal {\bf F}} (s)$ due to the lack of theoretical predictions. On the other hand, the theoretical problem is well-posed. When the exchange and dipolar coupling strength within a pair are smaller than $\delta_{\s 0}$ (which we assume), the shape of the Fourier transform depends only on two parameters: $\delta_{\s 0}$, which is the r.m.s. value of $\delta$, and $\Omega_{\s R}$.

In the following, the Fourier components of the Rabi spectrum ${\mathcal {\bf F}} (s)$ are calculated analytically. We find that the width of the crossover region is broad and extends from $\frac{\Omega_{\s R}}{\delta_{\s 0}}\approx 0.3$, where the $s=2\Omega_{\s R}$ peak appears, to $\frac{\Omega_{\s R}}{\delta_{\s 0}}\approx 2$, where it dominates over $s=\Omega_{\s R}$ peak. Most importantly, the analytical treatment reveals that ${\mathcal {\bf F}} (s)$ consists not only of two peaks (as discussed in most experimental studies) but rather of {\em three} peaks. The origin of the third peak, which  occurs at frequencies $s\ll \Omega_{\s R}$, is due a disorder induced distribution of $\delta$, which implies that even at strong $\Omega_{\s R}$, the two spins in the pair do not precess entirely in phase. This third peak is harder to observe experimentally compared to the two with higher frequencies, yet a prediction of its shape and evolution with $\Omega_{\s R}$ is provided. Also, a correlation study of disorder within pairs affects the shape of an ensemble average of ${\mathcal {\bf F}} (s)$. This follows from previous numerical study of the ratio between the magnitudes of the $s = \Omega_{\s R}$ and $s = 2\Omega_{\s R}$ Rabi oscillation peaks~\cite{Boehme3} which was conducted for particular value sets  of $\delta_{\s 0}$ and  $\Omega_{\s R}$.

The study presented here is outlined as follows. In Sect. II, a qualitative derivation of
conductivity changes in pEDMR experiments is given which reproduces
the result of rigorous consideration in Ref.~\onlinecite{BoehmeMain}.
In Sect. III, the analytical expression for a disorder averaged Fourier transform of $\Delta \sigma (\tau)$ is presented before this is used in Sect. IV for the consideration of correlation effects between the two random distribution of the two pair partner resonances. An analysis and discussion of these results is then presented in Sect. V.

%%%%%%%%%%%%%%%%%%%%%%%%%%%%%%%%%%%%%%%%%%%%%%%%%%%%%%%%%%%%%%%%%%%%%%%%%%%%%%%%%%%%%%%%%%%%%%%%%%%%%%%%%%%%%%%%%%%%%%%%%%%%%%%%%%%%%%%%%%%%%%%%%%%%%%%%%%%%%%%%%%%%%%%%%%%%%%%%%%%%%%%%%%%%%%% %%%%%%%%%%%%%%%%%%%%%%%%%%%%%%%%%%%%%%%%%%%%%%%%%%%%%%%%%%%%%%%%%%%%%%%%%%%%%%%%%%%%%%%%%%%%%%%%%%%%%%%%%%%%%%%%%%%%%%%%%%%%%%%%%%%%%%%%%%%%%%%%%%%%%%%%%%%%%%%%%%%%%%%%%%%%%%%%%%%%%%%%%%%%%%% %%%%%%%%%%%%%%%%%%%%%%%%%%%%%%%%%%%%%%%%%%%%%%%%%%%%%%%%%%%%%%%%%%%%%%%%%%%%%%%%%%%%%%%%%%%%%%%%%%%%%%%%%%%%%%%%%%%%%%%%%%%%%%%%%%%%%%%%%%%%%%%%%%%%%%%%%%%%%%%%%%%%%%%%%%%%%%%%%%%%%%%%%%%%%%% %%%%%%%%%%%%%%%%%%%%%%%%%%%%%%%%%%%%%%%%%%%%%%%%%%%%%%%%%%%%%%%%%%%%%%%%%%%%%%%%%%%%%%%%%%%%%%%%%%%%%%%%%%%%%%%%%%%%%%%%%%%%%%%%%%%%%%%%%%%%%%%%%%%%%%%%%%%%%%%%%%%%%%%%%%%%%%%%%%%%%%%%%%%%%%%
\section{The dependence of pEDMR induced conductivity changes on the pulse duration $\tau$}
The pair partners within a spin pair are denoted with $a$ and $b$ respectively. Before a pulse is applied to a spin pair, it will rest in one of its four spin--eigenstates, as both $a$ and $b$  can be either in a $\vert \!\downarrow \rangle$ or in  a $|\!\uparrow\rangle$ state. The important qualitative observation made in Ref. \onlinecite{Boehme0} is that only initial configurations $|\!\downarrow\downarrow\rangle$ and $|\!\uparrow \uparrow\rangle$ of the pair exist as the other two eigenstates with singlet content are very short lived. The  $|\!\downarrow\downarrow\rangle$ and $|\!\uparrow \uparrow\rangle$ states are therefore responsible for the change of conductivity after the end of the pulse. Without confinement of generality, one can assume that at $t=0$ both $a$ and $b$ are in the $\vert \!\downarrow \rangle$ state. The respective populations of the  $\vert\! \downarrow \rangle$ states will then evolve according to the Rabi formula
\be
\label{nab}
n_{a,b}(t)=1-\frac{\Omega_{\s R}^2}{\Omega_{\s R}^2+\delta_{\s a,b}^2}\sin^2
\sqrt{\frac{1}{4}\Big(\Omega_{\s R}^2+\delta_{\s a,b}^2\Big)}t,
\ee
where $\delta_a=\omega_a-\omega$ and $\delta_b=\omega_b-\omega$ are, the detuning frequencies of $a$ and $b$, respectively. A detuning frequency is the difference of a Larmor frequency $\omega_a$ or $\omega_b$ and the frequency $\omega$ of the driving field. After the pulse ends at $t=\tau$ the pair is in the $|\!\downarrow\downarrow\rangle$ state with probability $P_{\downarrow\downarrow}=n_a(\tau)n_b(\tau)$ and in $|\!\uparrow\uparrow\rangle$ state with probability $P_{\uparrow\uparrow}=\bigl(1-n_a(\tau)\bigr)\bigl(1-n_b(\tau)\bigr)$.
Then the  probability to find the pair in one of the states $|\!\downarrow\downarrow\rangle$ or $|\!\uparrow\uparrow\rangle$ is equal to
\begin{align}
\label{P1}
&P(\tau)=P_{\uparrow\uparrow}+P_{\downarrow\downarrow}&
\nonumber \\
&\hspace{0.7cm}=1-n_{\s a}(\tau)-n_{\s b}(\tau)+2n_{\s a}(\tau)n_{\s b}(\tau).&
\end{align}
It is easy to see that Eq.~\eqref{P1} also applies when the pair is initially in $|\!\uparrow\uparrow\rangle$ state. Eq. (\ref{P1}) coincides with the corresponding expression for the $\tau$-dependent part of the diagonal elements of the density matrix of the pair derived in Ref.~\onlinecite{BoehmeMain}. The probability $P(\tau)$ serves as the initial condition for the transient restoration of the steady state current after the pulse\cite{BoehmeMain}. Thus, $\Delta\sigma$ can be identified with $P(\tau)$ within a factor. An expression for $\Delta\sigma(\tau)$ averaged over the contributions from all pairs within a pair ensemble is then obtained from
\be
\label{average}
\big\langle \Delta \sigma(\tau) \big\rangle =\frac{1}{2\pi\delta_{\s 0}^2}
\int d\delta_{\s a}  d \delta_{\s b}
\exp\left[-\frac{\delta_{\s a}^2+\delta_{\s b}^2}{2 \delta_{\s 0}^2}\right]
\Delta\sigma(\delta_{\s a}, \delta_{\s b}, \tau).
\ee
Note that this expression for $\big\langle \Delta \sigma(\tau)\big\rangle$ can also be written as
%\be
%\label{T}
%\big\langle \Delta \sigma(\tau) \big\rangle =1-2 T(t)+2T^2(t),
%\ee
\be
\label{T}
\big\langle \Delta \sigma(\tau) \big\rangle =1-2 T(\tau)+2T^2(\tau),
\ee
where the function $T(\tau)$ is defined as
%\begin{align}
%T(t)=& \frac{1}{\sqrt{2\pi}\delta_{\s 0}}\int d\delta\,\, e^{-\delta^2/2\delta_{\s 0}^2}
%\left(\frac{\Omega_{\s R}^2}{\Omega_{\s R}^2+\delta^2}\right)
%\nonumber \\
%&\times
%\sin^2\sqrt{\frac{1}{4}\Big(\Omega_{\s R}^2+\delta^2\Big)}\,t.
%\end{align}
\begin{align}
T(\tau)=& \frac{1}{\sqrt{2\pi}\delta_{\s 0}}\int d\delta\,\, e^{-\delta^2/2\delta_{\s 0}^2}
\left(\frac{\Omega_{\s R}^2}{\Omega_{\s R}^2+\delta^2}\right)
\nonumber \\
&\times
\sin^2\sqrt{\frac{1}{4}\Big(\Omega_{\s R}^2+\delta^2\Big)}\,\tau.
\end{align}
In the limit of  long pulses $T(\tau)$ approaches a constant in an oscillatory fashion; the amplitude of the oscillations falls off slowly, as $\tau^{-1/2}$, with the length of the pulse~\cite{Loss}. For strong disorder ($\delta\gg\Omega_R$) the derivative can be expressed through the zero-order Bessel function~\cite{BoehmeNature} $T^{\prime}(\tau)=2^{-3/2}\pi^{1/2}\Omega_{\s R}^3\delta_{\s 0}^{-1}J_{\s 0}(\Omega_{\s R}\tau)$.

While the second term in Eq. (\ref{T}) describes Rabi oscillations within either component $a$ or $b$ of the pair, the third term ``knows" about the collective spin precession of $a$ and $b$. However, the $T^2$-term also contains contributions from the individual precessions of pair partners $a$ and $b$. We will therefore subtract these contributions and group them with the $T$-term in Eq. (\ref{T}) prior to performing the Fourier transform. By substituting  Eq. (\ref{nab}) into Eq. (\ref{P1}), we get
\begin{align}
\label{sigmafin}
&\Delta\sigma(\tau)=\frac{1}{2}+\frac{
\delta_{\s a}^2 \delta_{\s b}^2}{2(\Omega_{\s R}^2+\delta_{\s a}^2)(\Omega_{\s R}^2+\delta_{\s b}^2)}+\Big[G_{\s 1}(\delta_{\s a},\delta_{\s b},\tau)
\nonumber \\
&\hspace{1cm}
+G_{\s 1}(\delta_{\s b},\delta_{\s a},\tau)\Big]
+G_{\s -}(\delta_{\s a},\delta_{\s b},\tau)
+G_{\s +}(\delta_{\s a},\delta_{\s b},\tau),&
\end{align}
where the functions describing the three harmonic Rabi oscillation peaks are defined as
\be
\label{G1}
G_{\s 1}(\delta_{\s a},\delta_{\s b},\tau)=\frac{\Omega{\s R}^2\delta_{\s b}^2}{2}
\left[
\frac{\cos\big( \sqrt{\Omega_{\s R}^2 +\delta_{\s a}^2}\, \tau\ \big)
}{(\Omega_{\s R}^2+\delta_{\s a}^2)(\Omega_{\s R}^2+\delta_{\s b}^2)}\right]\!\!,
\ee
\be
\label{G-}
G_{\s -}(\delta_{\s a},\delta_{\s b},\tau)=\frac{\Omega{\s R}^4}{4}\!\!
\left[
\frac{\cos\Big\{\big( \sqrt{\Omega_{\s R}^2 +\delta_{\s a}^2}-\sqrt{\Omega_{\s R}^2 +\delta_{\s b}^2}\,\big) \tau\! \Big\}}{(\Omega_{\s R}^2+\delta_{\s a}^2)(\Omega_{\s R}^2+\delta_{\s b}^2)}\right]\!\!,
\ee
\be
\label{G+}
G_{\s +}(\delta_{\s a},\delta_{\s b},\tau)=\frac{\Omega{\s R}^4}{4}\!\!
\left[\frac{\cos\Big\{\big( \sqrt{\Omega_{\s R}^2 +\delta_{\s a}^2}+\sqrt{\Omega_{\s R}^2 +\delta_{\s b}^2}\,\big) \tau\! \Big\}}{(\Omega_{\s R}^2+\delta_{\s a}^2)(\Omega_{\s R}^2+\delta_{\s b}^2)}\right]\!\!.
\ee
The above terms $G_{\s 1}$, $G_{\s +}$, $G_{\s -}$ describe the peaks $s=\Omega_{\s R}$, $s=2\Omega_{\s R}$, and $s \ll \Omega_{\s R}$, contained in the Fourier transform $\bf{\mathrm{F}}(s)$, respectively.

%%%%%%%%%%%%%%%%%%%%%%%%%%%%%%%%%%%%%%%%%%%%%%%%%%%%%%%%%%%%%%%%%%%%%%%%%%%%%%%%%%%%%%%%%%%%%%%%%%%%%%%%%%%%%%%%%%%%%%%%%%%%%%%%%%%%%%%%%%%%%%%%%%%%%%%%%%%%%%%%%%%%%%%%%%%%%%%%%%%%%%%%%%%%%%% %%%%%%%%%%%%%%%%%%%%%%%%%%%%%%%%%%%%%%%%%%%%%%%%%%%%%%%%%%%%%%%%%%%%%%%%%%%%%%%%%%%%%%%%%%%%%%%%%%%%%%%%%%%%%%%%%%%%%%%%%%%%%%%%%%%%%%%%%%%%%%%%%%%%%%%%%%%%%%%%%%%%%%%%%%%%%%%%%%%%%%%%%%%%%%% %%%%%%%%%%%%%%%%%%%%%%%%%%%%%%%%%%%%%%%%%%%%%%%%%%%%%%%%%%%%%%%%%%%%%%%%%%%%%%%%%%%%%%%%%%%%%%%%%%%%%%%%%%%%%%%%%%%%%%%%%%%%%%%%%%%%%%%%%%%%%%%%%%%%%%%%%%%%%%%%%%%%%%%%%%%%%%%%%%%%%%%%%%%%%%% %%%%%%%%%%%%%%%%%%%%%%%%%%%%%%%%%%%%%%%%%%%%%%%%%%%%%%%%%%%%%%%%%%%%%%%%%%%%%%%%%%%%%%%%%%%%%%%%%%%%%%%%%%%%%%%%%%%%%%%%%%%%%%%%%%%%%%%%%%%%%%%%%%%%%%%%%%%%%%%%%%%%%%%%%%%%%%%%%%%%%%%%%%%%%%%
\section{Averaging over disorder within a spin pair ensemble}
Variations of the magnetic resonance frequency of each pair partner in each individual pair can occur due to: (i) variations of the spin--orbit coupling which changes the $g$--factor~\cite{spaeth2003point}. This is seen in disordered materials where the lengths and angles of chemical bonds can vary strongly~\cite{PhysRevB.79.195205, lee_192104}. (ii) Due to random hyperfine fields which can strongly fluctuate throughout a material because of the small polarization nuclear spins even at low temperature and high magnetic fields~\cite{BoehmeDifferentiation, nguyen2010isotope}. We define the Fourier spectrum of the conductivity change from the steady state as
\begin{align}
\label{Fourier}
&{\bf\mathrm{F}}(s)=\int_0^\infty d\tau\cos(s\tau)
\Big\langle \big(\Delta\sigma(\tau)-\Delta\sigma(0)\big)\Big\rangle .&
\end{align}
The expression can be decomposed into three contributions ${\bf\mathrm{F}}(s)={\bf\mathrm{F}}_{\s 1}(s)+{\bf\mathrm{F}}_{\s 0}(s)+{\bf\mathrm{F}}_{\s 2}(s)$, which derive from the terms $G_{\s 1}$, $G_{\s {-}}$, and $G_{\s {+}}$ terms in Eq. \eqref{sigmafin}. Obviously, the time integration of each term yields a $\delta$-function.

Our task is to perform the averaging of each $\delta$-function over disorder, as in Eq. \eqref{average}. We start from ${\bf\mathrm{F}}_1(s)$, which describes a peak near $s=\Omega_{\s R}$. For this contribution, the averaging over $\delta_a$, $\delta_b$ reduces to the product of averages
\begin{flalign}
&{\bf \mathrm{F}}_{\s 1}(s)=\frac{\Omega_{\s R}^2}{8\delta_{\s 0}^2}\int d\delta_{\s a} \,e^{-\delta_{\s a}^2/2\delta_{\s 0}^2}\,
\left[\frac{\delta(s-\sqrt{\Omega_{\s R}^2+\delta_{\s a}^2})}{\Omega_{\s R}^2+\delta_{\s a}^2}\right]&
\nonumber \\
&\hspace{1.3cm}\times \int d\delta_{\s b}\,\, e^{-\delta_{\s b}^2/\delta_{\s 0}^2}\left(\frac{ \delta_{\s b}^2}{\Omega_{\s R}^2+\delta_{\s b}^2}\right)\!.&
\end{flalign}
It is convenient to evaluate the integral over $\delta_{\s a}$ with the help of the $\delta$-function. The second integral can be reduced to the error-function leading to
\begin{flalign}
\label{F1 nocorr}
&{\bf \mathrm{F}}_{\s 1}(s)=\left(\frac{\Omega_{\s R}^3}{4\delta_{\s 0}^2s\sqrt{s^2-\Omega_{\s R}^2}}\right)\exp\left[-\frac{s^2-\Omega_{\s R}^2}{2\delta_{\s 0}^2}\right]
f\left(\frac{\Omega_{\s R}^2}{2\delta_{\s 0}^2}\right)\!,&
\end{flalign}
where the function $f$ is defined as
\begin{eqnarray}
f(b)&=&\int_{-\infty}^{\infty}dy\,e^{-y^2 b}\,\left( \frac{y^2}{1+y^2}\right)
\nonumber \\
&=& \sqrt{\pi}\left[\frac{1}{b^{1/2}}-\sqrt{\pi}e^b \,\,\mathrm{erfc}(\sqrt{b})
%\int_{\sqrt{b}}^\infty dv \,e^{-v^2}
\right]\!.
\end{eqnarray}
For ${\bf\mathrm{F}}_2(s)$, it is convenient, after substituting \eqref{G+}  into \eqref{Fourier}, to perform an integration over $\delta_a$, $\delta_b$ in polar coordinates. Upon introducing the new variables
\be
\label{polar}
\delta_{\s a}= \sqrt{2}v \cos \phi,~ \delta_{\s b}= \sqrt{2}v \sin \phi,
\ee
the expression for ${\bf\mathrm{F}}_2(s)$ acquires the form
\begin{flalign}
\label{def L2}
&{\bf \mathrm{F}}_{\s 2}(s)=\frac{\Omega_{\s R}^4}{8\delta_{\s 0}^2} \int_0^\infty dv \,v\, e^{-v^2/\delta_{\s 0}^2}
\int_0^{2\pi} d \phi&
\nonumber \\ &\hspace{0.4cm}\times
\frac{\delta\Big(s-\sqrt{\Omega_{\s R}^2 +2v^2 \cos^2 \phi}-\sqrt{\Omega_{\s R}^2 +2v^2 \sin^2 \phi}\Big)}
{\big(\Omega_{\s R}^2 +2v^2 \cos^2 \phi\big)\big(\Omega_{\s R}^2 +2v^2 \sin^2 \phi\big)}.
%{(\Omega_{\s R}^4+2\Omega_{\s R}^2v^2+4v^4\cos^2\phi \sin^2\phi)}.&%\hspace{2cm}.
\end{flalign}
Without the denominator, it is straightforward to perform an integration over $\phi$, which yields
\begin{eqnarray}
\label{angint}
&&\!\!\!\int_0^{2\pi} \!\!d \phi \,\delta\Big(s-\sqrt{\Omega_{\s R}^2 +2v^2 \cos^2 \phi}-\sqrt{\Omega_{\s R}^2 +2v^2 \sin^2 \phi}\Big)
\nonumber \\
&& \hspace{0.5cm}=\frac{2|2\Omega_{\s R}^2+2v^2-s^2|}
{\sqrt{\left[\left(\frac{s^2}{2}-v^2\right)^2-\Omega_{\s R}^2s^2\right]\Big[(\Omega_{\s R}^2+v^2)-\frac{s^2}{4}\Big]}}.
\end{eqnarray}
With the denominator, Eq. (\ref{angint}) is to be divided by the value $\frac{1}{4}(s^2-2\Omega_{\s R}^2-2v^2)^2$ of the denominator
where the argument of the $\delta$-function is zero, which yields
\begin{flalign}
\label{L2exact}
&{\bf \mathrm{F}}_{\s 2}(s)=
\frac{\Omega_{\s R}^4}{\delta_{\s 0}^2}
\bigintsss_0^{\sqrt{\frac{s^2}{2}-s\Omega_{\s R}}}
\frac{dv \,v\, e^{-v^2/\delta_{\s 0}^2}}{\big|s^2-2\Omega_{\s R}^2-2v^2\big|}&
\nonumber \\ &\hspace{0.8cm}\times
\frac{1}{\sqrt{\left[\left(\frac{s^2}{2}-v^2\right)^2-\Omega_{\s R}^2s^2\right]\left[(\Omega_{\s R}^2+v^2)-\frac{s^2}{4}\right]}}.
&
\end{flalign}
Eq.~\eqref{L2exact} is defined only for $s>2\Omega_{\s R}$. As $s$ approaches $2\Omega_{\s R}$,  both brackets under the square root turn to zero. At the same time, the integration interval also shrinks to zero. However, the factor  $|s^2-2\Omega_{\s R}^2-2v^2|$ in the denominator is nonsingular near $s=2\Omega_{\s R}$. To illuminate the behavior of ${\bf\mathrm{F}}_2(s)$ near the threshold, it is convenient to make the  substitution $w=v^2-\frac{s^2}{4}+\Omega_{\s R}^2$ in the integral of Eq. \eqref{L2exact}. It then assumes the form
\begin{flalign}
\label{F2exact}
&{\bf \mathrm{F}}_{\s 2}(s)=\left(\frac{\Omega_{\s R}^4}{2\delta_{\s 0}^2}\right)
e^{-\frac{\frac{s^2}{4}-\Omega_{\s R}^2}{\delta_{\s 0}^2}}\bigintsss\limits_0^{(\frac{s}{2}-\Omega_{\s R})^2}
\hspace{-0.2cm}\frac{dw\,e^{-w/\delta_{\s 0}^2}}{\sqrt{w\left[\left(\frac{s}{2}-\Omega_{\s R}\right)^2-w\right]}}&
\nonumber \\
&\hspace{1.1cm}\times
\frac{1}{|\frac{s^2}{2}-2w|\sqrt{
\left[\left(\frac{s}{2}+\Omega_{\s R}\right)^2-w\right]}}.&
\end{flalign}
Now we see that only the first two factors in the denominator are singular when $s$ is close to $2\Omega_{\s R}$, where the $s=2\Omega_{\s R}$ peak occurs. In this domain we can set $w=0$ in the last two factors and take them out of the integrand. Then, the remaining integral readily reduces to the modified Bessel function,
$I_{\s 0}(y)$, and we get
\be
\label{SecondPeak}
{\bf \mathrm{F}}_{\s 2}(s)=
\left(\frac{2\Omega_{\s R}^4}{\delta_{\s 0}^2s^2}\right)
\left(\frac{e^{-\frac{s^2/4-\Omega_{\s R}^2}{\delta_{\s 0}^2}}}
{s+2\Omega_{\s R}}\right)
G\left[\frac{\left(s-2\Omega_{\s R}\right)^2}{4\delta_{\s 0}^2}\right]\!,
\ee
where
\be
\label{H}
G(b)=\bigintsss_0^{b}\frac{dx \,\,e^{-x}}{\sqrt{x(b-x)}}
= \pi e^{-b/2}\,\,I_{\s 0}\left(\frac{b}{2}\right).
\ee
The analysis of the shape of the $s=2\Omega_{\s R}$ peak given in Sect. V. will reveal that the approximation in Eq. (\ref{SecondPeak}) describes not only the vicinity  $(s-2\Omega_{\s R}) \ll \Omega_{\s R}$ but the entire peak when $\Omega_{\s R}$ is bigger than $0.3\delta_{\s 0}$.

Finally we turn our attention to the peak at $s \ll \Omega_{\s R}$. The initial expression for  ${\bf \mathrm{F}}_{\s 2}(s)$ differs from Eq. \eqref{def L2} for ${\bf \mathrm{F}}_{\s 0}(s)$ only in one respect. Instead of the sum, $\sqrt{\Omega_{\s R}^2 +\delta_{\s a}^2}+\sqrt{\Omega_{\s R}^2 +\delta_{\s a}^2}$, in the argument of the $\delta$-function it contains a difference, $\sqrt{\Omega_{\s R}^2 +\delta_{\s a}^2}-\sqrt{\Omega_{\s R}^2 +\delta_{\s a}^2}$. The angular integration in polar coordinates is therefore performed in a similar way as Eq.~\eqref{angint}
\begin{eqnarray}
\label{angint2}
&&\!\!\!\int_0^{2\pi} \!\!d \phi \,\delta\Big(s-\sqrt{\Omega_{\s R}^2 +2v^2 \cos^2 \phi}+\sqrt{\Omega_{\s R}^2 +2v^2 \sin^2 \phi}\Big)
\nonumber \\
&&\hspace{0.3cm}= \frac{|2\Omega_{\s R}^2+2v^2-s^2|}
{\sqrt{\left[\left(\frac{s^2}{2}-v^2\right)^2-\Omega_{\s R}^2s^2\right]\Big[(\Omega_{\s R}^2+v^2)-\frac{s^2}{4}\Big]}}.
\end{eqnarray}
Note that, compared to Eq. \eqref{angint2}, the integral Eq. \eqref{angint} has an extra factor of $2$. This is because the argument of $\delta$-function in Eq. \eqref{angint2} turns to zero  at two values of $\phi$, while in Eq. \eqref{angint}, it turns to zero at four values of $\phi$. To get the final expression for ${\bf \mathrm{F}}_{\s 0}$ one has again to divide Eq. \eqref{angint2} by the value of denominator at the zeros points of the $\delta$-function, which is equal to $\frac{1}{4}(s^2-2\Omega_{\s R}^2-2v^2)^2$, i.e., the same as in ${\bf \mathrm{F}}_{\s 2}(s)$. This leads to
\begin{flalign}
&{\bf \mathrm{F}}_{\s 0}(s)=
\frac{\Omega_{\s R}^4}{2\delta_{\s 0}^2}
\bigintsss\limits_{\sqrt{\frac{s^2}{2}+s\Omega_{\s R}}}^\infty
\frac{dv \,v\, e^{-v^2/\delta_{\s 0}^2}}
{\sqrt{\left(\frac{s^2}{2}-v^2\right)^2-\Omega_{\s R}^2s^2}}&
\nonumber \\ &\hspace{1.1cm}\times
\frac{1}{\big|s^2-2\Omega_{\s R}^2-2v^2\big|\sqrt{(\Omega_{\s R}^2+v^2)-\frac{s^2}{4}}}.
&
\end{flalign}
We see again that only the first factor in the denominator is singular at the lower limit $v=\bigl(\frac{s^2}{2}+s\Omega_{\s R}\bigr)^{1/2}$. Thus, for small $s$, the peak is described
by substituting $s=0$ into the second and third factors in denominator, which leads to
\be
{\bf \mathrm{F}}_{\s 0}(s)=
\frac{\Omega_{\s R}}{4\delta_{\s 0}^2}
\bigintss_{\sqrt{{\frac{s^2}{2}+s\Omega_{\s R}}}}^\infty
\frac{dv\,v\, e^{-v^2/\delta_{\s 0}^2}}{\sqrt{\left(\frac{s^2}{2}-v^2\right)^2-\Omega_{\s R}^2s^2}}.
\ee
The remaining integral can be expressed via the Macdonald function, $K_{\s 0}(y)$, which yields
\be
\label{smallS}
{\bf \mathrm{F}}_{\s 0}(s)=
\left(\frac{\Omega_{\s R}}{8\delta_{\s 0}^2}\right)
e^{-s^2/2\delta_{\s 0}^2}
\,\,K_{\s 0}\left(\frac{s\Omega_{\s R}}{\delta_{\s 0}^2}\right).
\ee
We will see in Sect. V. that the expression in Eq. (\ref{smallS}) describes the entire peak when $\Omega_{\s R}$ is big enough, $\Omega_{\s R} \gtrsim \delta_{\s 0}$.

%%%%%%%%%%%%%%%%%%%%%%%%%%%%%%%%%%%%%%%%%%%%%%%%%%%%%%%%%%%%%%%%%%%%%%%%%%%%%%%%%%%%%%%%%%%%%%%%%%%%%%%%%%%%%%%%%%%%%%%%%%%%%%%%%%%%%%%%%%%%%%%%%%%%%%%%%%%%%%%%%%%%%%%%%%%%%%%%%%%%%%%%%%%%%%% %%%%%%%%%%%%%%%%%%%%%%%%%%%%%%%%%%%%%%%%%%%%%%%%%%%%%%%%%%%%%%%%%%%%%%%%%%%%%%%%%%%%%%%%%%%%%%%%%%%%%%%%%%%%%%%%%%%%%%%%%%%%%%%%%%%%%%%%%%%%%%%%%%%%%%%%%%%%%%%%%%%%%%%%%%%%%%%%%%%%%%%%%%%%%%% %%%%%%%%%%%%%%%%%%%%%%%%%%%%%%%%%%%%%%%%%%%%%%%%%%%%%%%%%%%%%%%%%%%%%%%%%%%%%%%%%%%%%%%%%%%%%%%%%%%%%%%%%%%%%%%%%%%%%%%%%%%%%%%%%%%%%%%%%%%%%%%%%%%%%%%%%%%%%%%%%%%%%%%%%%%%%%%%%%%%%%%%%%%%%%% %%%%%%%%%%%%%%%%%%%%%%%%%%%%%%%%%%%%%%%%%%%%%%%%%%%%%%%%%%%%%%%%%%%%%%%%%%%%%%%%%%%%%%%%%%%%%%%%%%%%%%%%%%%%%%%%%%%%%%%%%%%%%%%%%%%%%%%%%%%%%%%%%%%%%%%%%%%%%%%%%%%%%%%%%%%%%%%%%%%%%%%%%%%%%%%
\section{Correlation between the pair partner disorder}
Due to the proximity of the pair partners in each pair, the disorder related randomness of $\delta_{\s a}$ and $\delta_{\s b}$ may be correlated. Examples for such a correlation could be the common exposure of polaronic state in organic semiconductors to an overlapping nuclear spin bath or the correlation of the spin--orbit interaction in a disordered semiconductor due to local strain fields~\cite{PhysRevLett.106.037601}. In such cases, $\delta_{\s a}$ and $\delta_{\s b}$ are not statistically independent and the degree of overlap can be expressed by a correlation parameter, $x$ ($0\!<\!x\!<\!1$). Then the joint distribution function of  $\delta_{\s a}$, $\delta_{\s b}$ assumes the form
\be
\Phi(\delta_{\s a}, \delta_{\s b})= \frac{1}{2\pi \delta_{\s 0}^2\sqrt{1-x^2}}\exp{\left[-\frac{\delta_{\s a}^2+\delta_{\s b}^2-2x\delta_{\s a}^2\delta_{\s b}^2}{2\delta_{\s 0}^2(1-x^2)}\right]}\!.
\ee
We study the effect of correlation for the limit $\Omega_{\s R} \gg \delta_{\s 0}$ when all three peaks are well-developed and do not overlap. In this limit, we can use the expansion $\sqrt{\Omega_{\s R}^2+\delta_{{\s a},{\s b}}^2} \approx \Omega_{\s R}+\frac{\delta_{{\s a},{\s b}}^2}{2\Omega_{\s R}}$ in the arguments of the $\delta$-functions. We can also replace $\sqrt{\Omega_{\s R}^2+\delta_{{\s a},{\s b}}^2}$ by $\Omega_{\s R}$ in the denominators of Eqs. \eqref{G1} to \eqref{G+}. With these simplifications the expression for ${\bf \mathrm{F}}_{\s 1}(s)$ assumes the form
\begin{flalign}
&{\bf \mathrm{F}}_{\s 1}(s)=\frac{1}{8\delta_{\s 0}^2\Omega_{\s R}^2\sqrt{1-x^2}}\int\! d\delta_{\s b}\! \int d\delta_{\s a} \,\delta_{\s b}^2&
\nonumber \\
&\hspace{1.1cm}\times\exp\left[-\frac{\delta_{\s a}^2+\delta_{\s b}^2-2x \delta_{\s b} \delta_{\s a}}{2\delta_{\s 0}^2(1-x^2)}\right]
\delta \left(s-\Omega_{\s R}-\frac{\delta_{\s a}^2}{2\Omega_{\s R}}\right)\!.&
\end{flalign}
Integrating over $\delta_{\s a}$ with the help of the
$\delta$-function yields
\begin{flalign}
&{\bf \mathrm{F}}_{\s 1}(s)=
\frac{ \exp\left[-\frac{\Omega_{\s R}(s-\Omega_{\s R})}{\delta_{\s 0}^2(1-x^2)}\right]}
{\Big(4\Omega_{\s R}\delta_{\s 0}^2\sqrt{1-x^2}\Big)\sqrt{2\Omega_{\s R}(s-\Omega_{\s R})}}
\bigintsss\! d\delta_{\s b}&
\nonumber \\
&\times  \delta_{\s b}^2\,\exp\left[-\frac{\delta_{\s b}^2}{2\delta_{\s 0}^2(1-x^2)}\right]
\cosh\left(\frac{\delta_{\s b}x\sqrt{2\Omega_{\s R}(s-\Omega_{\s R})}}{\delta_{\s 0}^2(1-x^2)}\right)\!.&
\nonumber \\
\end{flalign}
Subsequent integration over $\delta_{\s b}$ is straightforward leading to
\begin{flalign}
\label{correlated1}
&{\bf \mathrm{F}}_{\s 1}(s)=\frac{\sqrt{\pi} \delta_{\s 0}(1-x^2)}{4\Omega_{\s R}\sqrt{\Omega_{\s R}(s-\Omega_{\s R})}}
\left[1+\frac{2x^2\Omega_{\s R}(s-\Omega_{\s R})}{\delta_{\s 0}^2(1-x^2)}\right]&
\nonumber \\
&\hspace{1.1cm}\times\exp\left[-\frac{\Omega_{\s R}(s-\Omega_{\s R})}{\delta_{\s 0}^2}\right]\!.&
\end{flalign}
We see that the correlation parameter, $x$, enters this expression only in the prefactor. In the limit of $x\rightarrow 0$, Eq. (\ref{correlated1}) matches Eq. \eqref{F1 nocorr} as can be seen when $\Omega_{\s R}\gg \delta_{\s 0}$ is assumed and Eq. \eqref{F1 nocorr} is expanded around $s=\Omega_{\s R}$.

In the limit of large $\Omega_{\s R}$ the definition of ${\bf \mathrm{F}}_{\s 2}(s)$ becomes
\begin{flalign}
\label{F2 defcorr}
&{\bf \mathrm{F}}_{\s 2}(s)=\left(\frac{1}{16\delta_{\s 0}^2\sqrt{1-x^2}}\right)
\int d\delta_{\s a} \,d\delta_{\s b}&
\nonumber \\
&\hspace{0.8cm}\times
\exp\left[-\frac{\delta_{\s a}^2+\delta_{\s b}^2-2x \delta_{\s b} \delta_{\s a}}{2\delta_{\s 0}^2(1-x^2)}\right]
%\exp\left[-\frac{(\delta_{\s a}+\delta_{\s b})^2}{2\delta_{\s 0}^2(1+x)}-\frac{(\delta_{\s a}-\delta_{\s b})^2}{4\delta_{\s 0}^2(1-x)}\right]
\delta \!\left(s-2\Omega_{\s R}-\frac{\delta_{\s a}^2+\delta_{\s b}^2}{2\Omega_{\s R}}\right)\!.&
%\frac{(\delta_{\s a}+\delta_{\s b})^2+(\delta_{\s a}-\delta_{\s b})^2}{4\Omega_{\s R}}\right)
\end{flalign}
Both integrals over $\delta_{\s a}$ and $\delta_{\s b}$ can be taken explicitly upon introduction of polar coordinates
\be
r=\sqrt{2(\delta_{\s a}^2+\delta_{\s b}^2)},
~\phi = \arctan\left(\frac{\delta_{\s a}-\delta_{\s b}}{\delta_{\s a}+\delta_{\s b}}\right),
\ee
so that Eq. \eqref{F2 defcorr} assumes the form
\begin{flalign}
&{\bf \mathrm{F}}_{\s 2}(s)=\left(\frac{1}{32\delta_{\s 0}^2\sqrt{1-x^2}}\right)
\int dr \,r \,d\phi&
\nonumber \\
&\hspace{1.1cm}\times\exp\left[-\frac{r^2(1+x\cos 2\phi)}{4\delta_{\s 0}^2(1-x^2)}\right]
%\nonumber \\
%&&\times
\delta \left(s-2\Omega_{\s R}-\frac{r^2}{4\Omega_{\s R}}\right)\!.&
\end{flalign}
After integrating over $r$ by using the $\delta$-function, the remaining integral over $\phi$ reduces to $I_{\s 0}(y)$, and we arrive at
\begin{align}
\label{F2 fincorr}
&{\bf \mathrm{F}}_{\s 2}(s)=\left(\frac{\pi\Omega_{\s R}}{8 \delta_{\s 0}^2\sqrt{1-x^2}}\right)
\exp\left[-\frac{\Omega_{\s R}(s-2\Omega_{\s R})}{\delta_{\s 0}^2(1-x^2)}\right]&
\nonumber \\
&\hspace{1.3cm}\times I_{\s 0}\left[\frac{\Omega_{\s R}x(s-2\Omega_{\s R})}{\delta_{\s 0}^2(1-x^2)}\right]\!.&
\end{align}
Similarly to the non--correlated case, ${\bf \mathrm{F}}_{\s 2}(s)$ is expressed through $I_{\s 0}(y)$. However, note that the argument of $I_{\s 0}(y)$ in Eq. \eqref{F2 fincorr} is completely different from Eq. \eqref{SecondPeak}. In fact, Eq. \eqref{F2 fincorr} was derived for the case when $I_{\s 0}\big(\frac{b}{2}\big)$ in Eq. \eqref{H} should be replaced by $1$.

In the presence of disorder correlation, the shape of ${\bf \mathrm{F}}_{\s 0}(s)$ can be also expressed via the Macdonald function with $x$-dependent argument. To see this, we take the definition
\begin{flalign}
\label{F0 corr}
&{\bf \mathrm{F}}_{\s 0}(s)=\left(\frac{1}{16\delta_{\s 0}^2\sqrt{1-x^2}}\right)
\int d\delta_{\s a} d\delta_{\s b}&
\nonumber \\
&\hspace{.3cm}\times
\exp\left[-\frac{\delta_{\s a}^2+\delta_{\s b}^2-2x \delta_{\s b} \delta_{\s a}}{2\delta_{\s 0}^2(1-x^2)}\right]
%\exp\left[-\frac{(\delta_{\s a}+\delta_{\s b})^2}{2\delta_{\s 0}^2(1+x)}-\frac{(\delta_{\s a}-\delta_{\s b})^2}{4\delta_{\s 0}^2(1-x)}\right]
\delta \left(s-\frac{\delta_{\s a}^2-\delta_{\s b}^2}{2\Omega_{\s R}}\right)\!,&
%\frac{(\delta_{\s a}+\delta_{\s b})^2+(\delta_{\s a}-\delta_{\s b})^2}{4\Omega_{\s R}}\right)
\nonumber \\
\end{flalign}
and introduce polar coordinates
\begin{align}
&r=\sqrt{\frac{\delta_{\s a}^2+\delta_{\s b}^2-2x \delta_{\s b} \delta_{\s a}}{2\delta_{\s 0}^2(1-x^2)}},&
\nonumber \\
&\phi=\arctan\left[
\sqrt{\frac{1+x}{1-x}}
\left(\frac
{\delta_{\s a}-\delta_{\s b}}{\delta_{\s a}+\delta_{\s b}}\right)
\right]\!.&
\end{align}
Eq.~\eqref{F0 corr} then becomes
\be
{\bf \mathrm{F}}_{\s 0}(s)=\frac{1}{8}
\int \!dr \,r\,e^{-r^2}  \!\!\!\int \!d\phi\,
\delta\left(s-\frac{\delta_{\s 0}^2r^2\sqrt{1-x^2}}{\Omega_{\s R}}\sin 2\phi\right)\!.
\ee
The integration over $\phi$ can be done explicitly by using the $\delta$-function, yielding
\be
{\bf \mathrm{F}}_{\s 0}(s)=\frac{1}{4s}
\bigintsss_{\left(\frac{s \Omega_{\s R}}{\delta_{\s 0}^2 \sqrt{1-x^2}}\right)^{1/2}}^\infty
\frac{dr\,r\,e^{-r^2} }{\sqrt{\frac{\delta_{\s 0}^4(1-x^2)}{\Omega_{\s R}^2s^2}r^4-1}}.
\ee
The reduction to the Macdonald function can then be achieved by the substitution $r^2=\frac{s \Omega_{\s R}}{\delta_{\s 0}^2 \sqrt{1-x^2}}r_{\s 1}$ which reveals
\be
\label{F0 fin}
{\bf \mathrm{F}}_{\s 0}(s)=
\left(\frac{\Omega_{\s R}}{8 \delta_{\s 0}^2\sqrt{1-x^2}}\right)
K_{\s 0}\left[\frac{s\Omega_{\s R}}{\delta_{\s 0}^2\sqrt{1-x^2}}\right]\!.
\ee

%%%%%%%%%%%%%%%%%%%%%%%%%%%%%%%%%%%%%%%%%%%%%%%%%%%%%%%%%%%%%%%%%%%%%%%%%%%%%%%%%%%%%%%%%%%%%%%%%%%%%%%%%%%%%%%%%%%%%%%%%%%%%%%%%%%%%%%%%%%%%%%%%%%%%%%%%%%%%%%%%%%%%%%%%%%%%%%%%%%%%%%%%%%%%%% %%%%%%%%%%%%%%%%%%%%%%%%%%%%%%%%%%%%%%%%%%%%%%%%%%%%%%%%%%%%%%%%%%%%%%%%%%%%%%%%%%%%%%%%%%%%%%%%%%%%%%%%%%%%%%%%%%%%%%%%%%%%%%%%%%%%%%%%%%%%%%%%%%%%%%%%%%%%%%%%%%%%%%%%%%%%%%%%%%%%%%%%%%%%%%% %%%%%%%%%%%%%%%%%%%%%%%%%%%%%%%%%%%%%%%%%%%%%%%%%%%%%%%%%%%%%%%%%%%%%%%%%%%%%%%%%%%%%%%%%%%%%%%%%%%%%%%%%%%%%%%%%%%%%%%%%%%%%%%%%%%%%%%%%%%%%%%%%%%%%%%%%%%%%%%%%%%%%%%%%%%%%%%%%%%%%%%%%%%%%%% %%%%%%%%%%%%%%%%%%%%%%%%%%%%%%%%%%%%%%%%%%%%%%%%%%%%%%%%%%%%%%%%%%%%%%%%%%%%%%%%%%%%%%%%%%%%%%%%%%%%%%%%%%%%%%%%%%%%%%%%%%%%%%%%%%%%%%%%%%%%%%%%%%%%%%%%%%%%%%%%%%%%%%%%%%%%%%%%%%%%%%%%%%%%%%%

\section{Analysis and discussion}
\begin{figure}[h!]
\includegraphics[width=80mm, angle=0,clip]{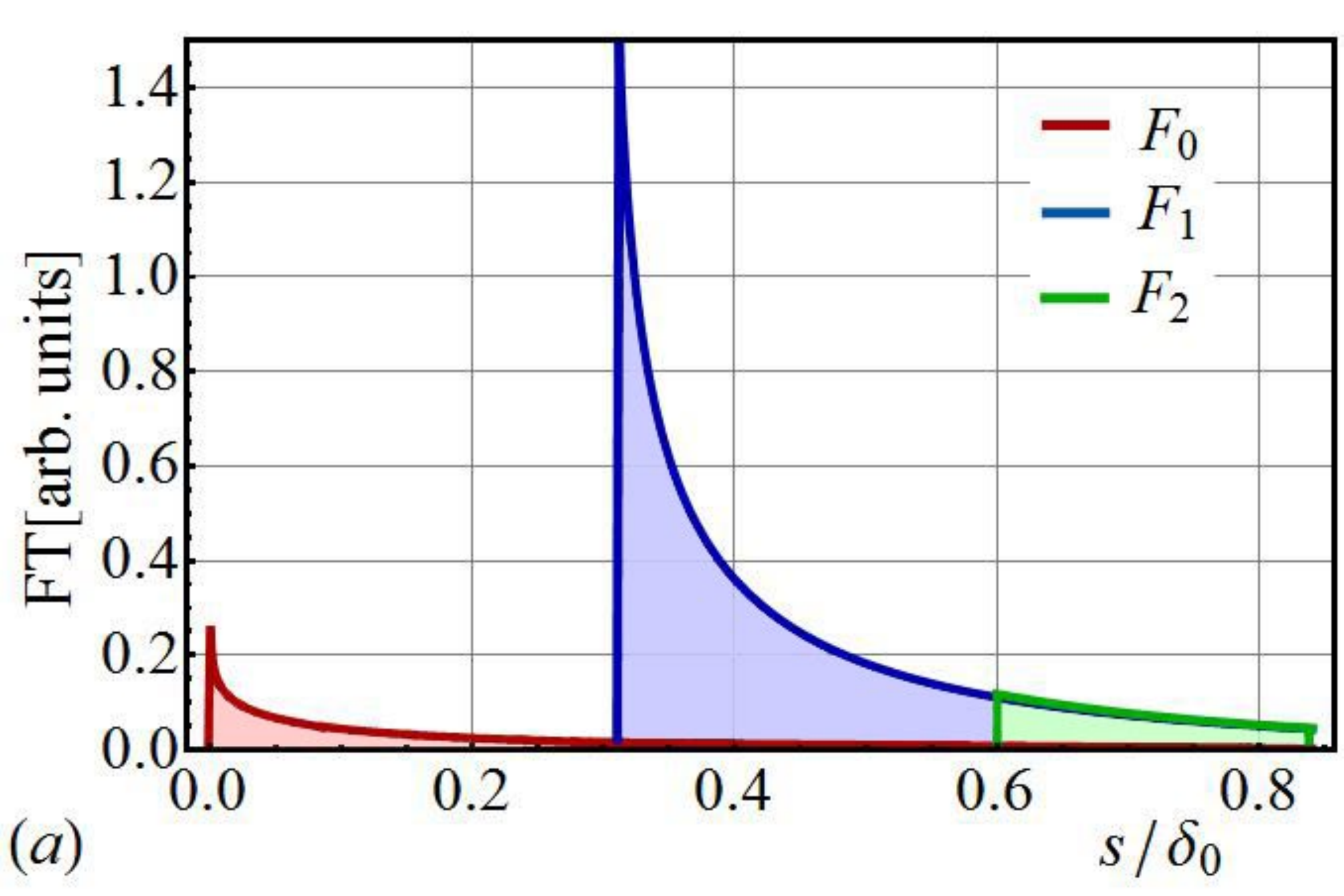}
\vspace{0.1cm}
\includegraphics[width=80mm, angle=0,clip]{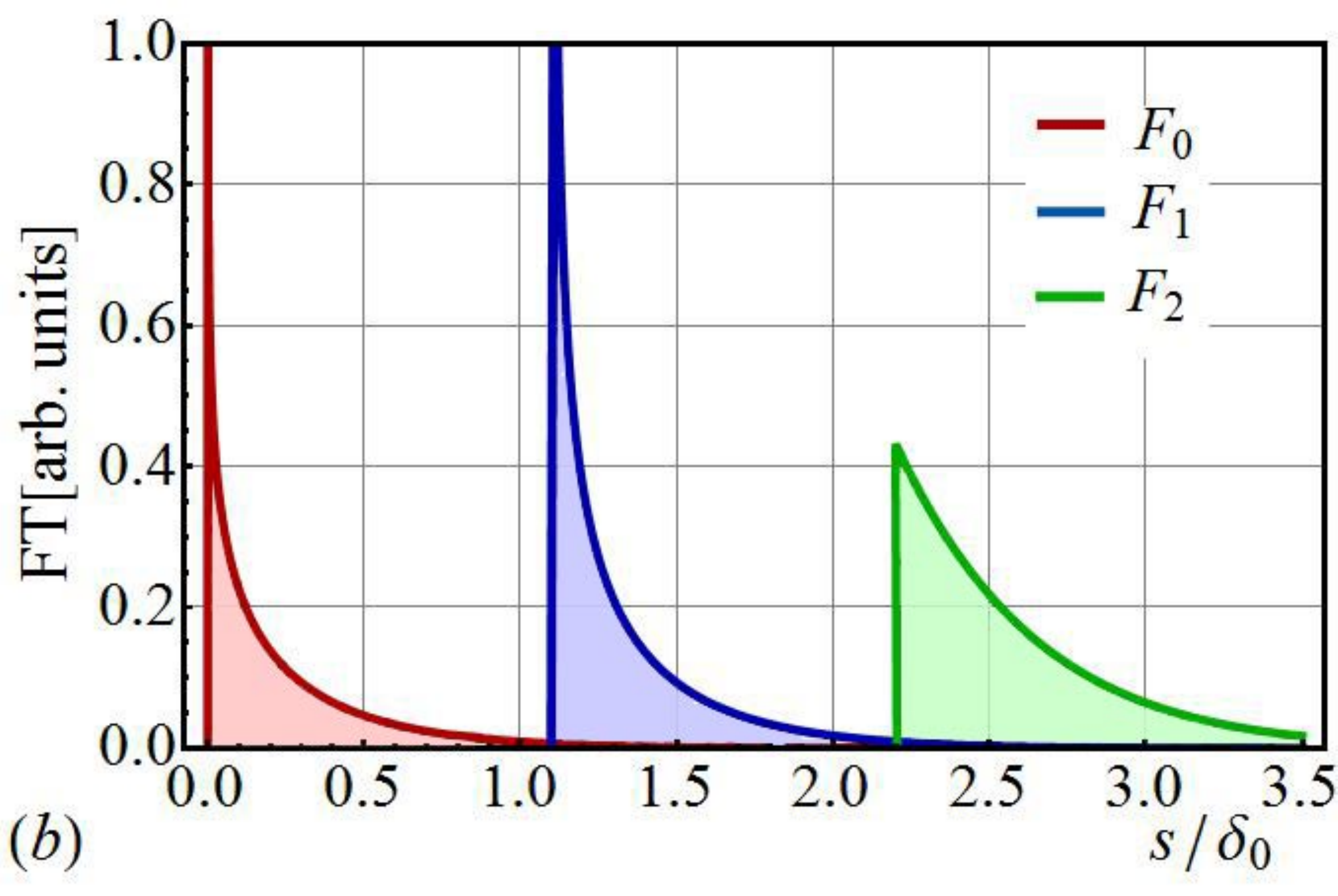}
\vspace{0.1cm}
\includegraphics[width=80mm, angle=0,clip]{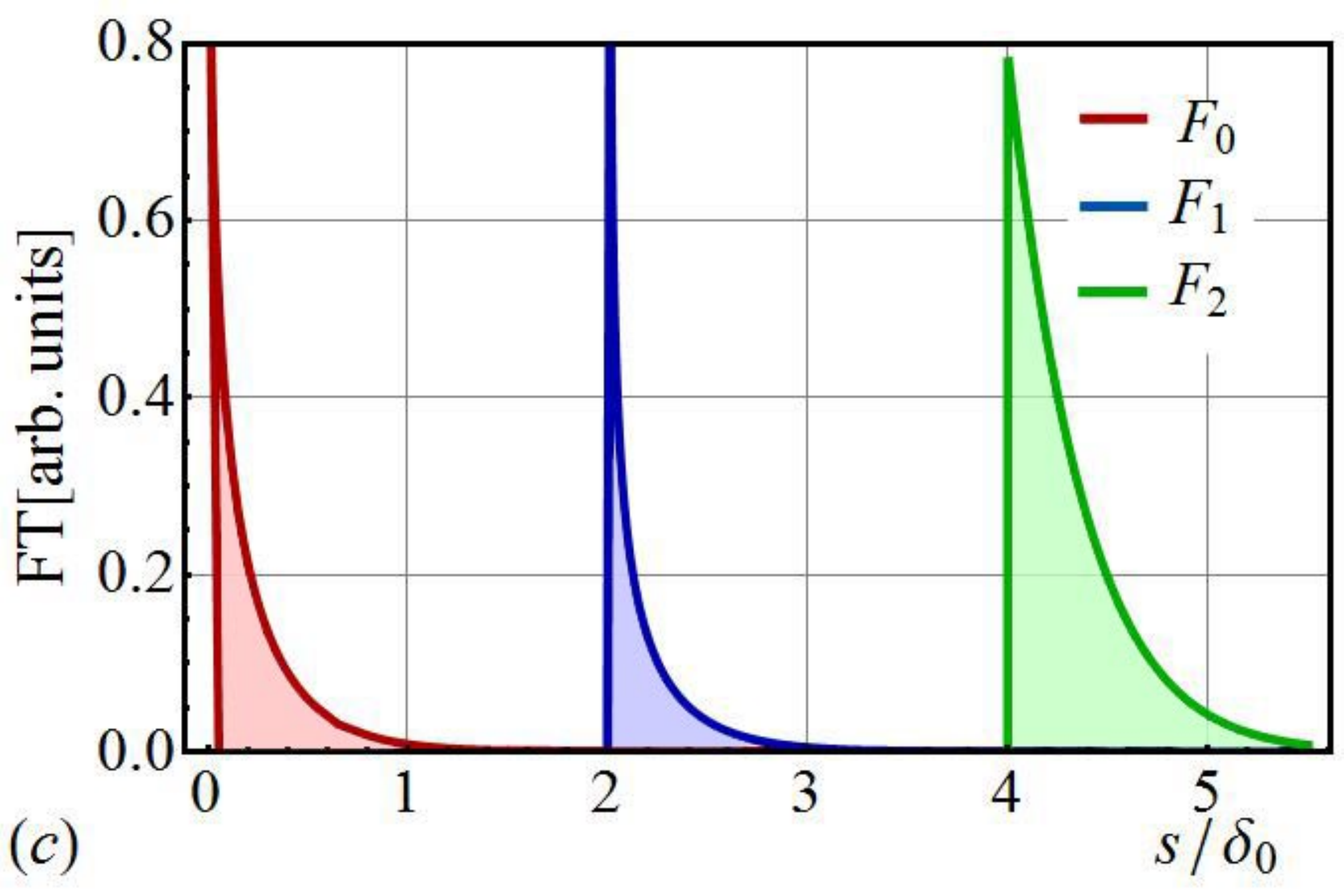}
\vspace{-0.4cm}
\caption{The shapes of the three peaks $s\ll\Omega_{\s R}$ (red), $s=\Omega_{\s R}$ (blue), and $s=2\Omega_{\s R}$ (green) in the Fourier transform of $\Delta \sigma(\tau)$ are plotted for three values of the dimensionless Rabi frequency $w=\Omega_{\s R}/\delta_{\s 0}$: (a) $w=0.3$, (b) $w=1.1$, (c) $w=2$. }
\label{fig:nocorr}
\end{figure}

The most important results of this study are the analytical expressions for the lineshapes of Rabi--oscillation peaks as they can be found in the Fourier analysis (the Rabi spectra) of the pulse length dependent conductivity changes $\Delta\sigma(\tau)$ that are measured with pEDMR experiments. For the case of uncorrelated disorder, Eqs. \eqref{F1 nocorr}, \eqref{SecondPeak}, and \eqref{smallS}, reveal these peak shapes for the oscillation peaks ${\bf \mathrm{F}}_{\s 1}(s)$,  ${\bf \mathrm{F}}_{\s 0}(s)$, and ${\bf \mathrm{F}}_{\s 2}(s)$, respectively. For the case of correlated disorder, the same peaks are described by Eqs. \eqref{correlated1}, \eqref{F2 fincorr}, and \eqref{F0 fin}.
We consider now the limit $\Omega_{\s R}\gg \delta_{\s 0}$, when all three peaks are well developed. It is easy to see that, for uncorrelated disorder, all three peaks exhibit the same exponential tail
\begin{eqnarray}
\label{tail}
{\bf \mathrm{F}}_{\s 1}(s)\approx\frac{\pi^{1/2}}{4\Omega_{\s R}}
\sqrt{\frac{{\s \Delta}s}{s-\Omega_{\s R}}}
e^{-\frac{s-\Omega_{\s R}}
{{\s \Delta} s}}\!,\nonumber\\
{\bf \mathrm{F}}_{\s 0}(s)\approx \frac{\sqrt{\pi}}{8\sqrt{2s{\s \Delta}s}}e^{-\frac{s}{{\s \Delta} s}}\!,~~
{\bf \mathrm{F}}_{\s 2}(s)\approx \frac{\pi}{8{{\s \Delta}} s}e^{-\frac{s-2\Omega_{\s R}}
{{\s \Delta} s}}\!,
\end{eqnarray}
where the characteristic width of the tail is given by
\begin{equation}
\label{width}
{\s \Delta}s=\frac{\delta_{\s 0}^2}{\Omega_{\s R}}.
\end{equation}
Naturally, all three peaks shrink with increasing $\Omega_{\s R}$. It is less trivial to realize  that, for the same deviation from the origin, the peaks ${\bf \mathrm{F}}_{\s 0}$, and ${\bf \mathrm{F}}_{\s 2}$ have a larger magnitude than ${\bf \mathrm{F}}_{\s 1}$, whose terms contain $\Omega_{\s R}\gg {\s \Delta}s$ in the denominator. This relation between the peaks is illustrated in Fig.~\ref{fig:nocorr}c.
%%%%%%%%%%%%%%%%%%%%%%%%%%%%%%%%%%%%%%%%%%%%%%%%%%%%%%%%%%%%%%%%%%%%%%%%%%%%%%%%%%%%%%%%%%%%%%%%%%%%%%%%%%%%%%%%%%%%%%%%%%%%%%%%%%%%%%%%%%%%%%%%%%%%%%%%%%%%%%%%%%%%%%%%%%%%%%%%%%%%%%%%%%%%%%%
%%%%%%%%%%%%%%%%%%%%%%%%%%%%%%%%%%%%%%%%%%%%%%%%%%%%%%%%%%%%%%%%%%%%%%%%%%%%%%%%%%%%%%%%%%%%%%%%%%%%%%%%%%%%%%%%%%%%%%%%%%%%%%%%%%%%%%%%%%%%%%%%%%%%%%%%%%%%%%%%%%%%%%%%%%%%%%%%%%%%%%%%%%%%%%%

We consider now the opposite limit of strong but uncorrelated disorder where $\delta_{\s 0}\gg \Omega_{\s R}$ and $x=0$. Eq. \eqref{SecondPeak} implies that ${\bf \mathrm{F}}_{\s 2}(2\Omega_{\s R})=\frac{\Omega_{\s R}}{16\delta_{\s 0}^2}$ is always finite at its peak. At the same time, the value of ${\bf \mathrm{F}}_{\s 1}(s)$ at $s=2\Omega_{\s R}$ is $\frac{\pi^{3/2}}{4\sqrt{6}\delta_{\s 0}}$, i.e., it is bigger than ${\bf \mathrm{F}}_{\s 2}(2\Omega_{\s R})$. Therefore, the $s=2\Omega_{\s R}$ peak is indistinguishable on the background of the $s=\Omega_{\s R}$ peak. This behavior reflects the physics of the weak driving regime~\cite{BoehmeMain} where only one component of the pair can be in resonance with the driving field at any time. Numerically, however, the  $s=2\Omega_{\s R}$ peak is pronounced already at $\Omega_{\s R}>0.3\delta_{\s 0}$, as shown in Fig. \ref{fig:nocorr}a. For the given strength of $\Omega_{\s R}$, the approximation Eq.~\eqref{SecondPeak} using the modified Bessel function is already justified.

The low frequency Rabi--oscillation peak which is described by Eq.~\eqref{smallS} diverges logarithmically in the limit of $s \rightarrow 0$ when ${\bf \mathrm{F}}_{\s 0}(s)\propto \ln(1/s)$. Nevertheless, the peak still loses to ${\bf \mathrm{F}}_{\s 1}(s)$ in the weak-driving regime due to its small prefactor $\frac{\Omega_{\s R}}{8 \delta_{\s 0}^2}$. The lineshape of this peak is described by Eq.~\eqref{smallS} when the ratio $\Omega_{\s R}/\delta{\s 0}$ exceeds $1$. This could be the reason why this peak has not received much attention in previous experimental studies since this regime is hard (yet not impossible) to attain experimentally.
%%%%%%%%%%%%%%%%%%%%%%%%%%%%%%%%%%%%%%%%%%%%%%%%%%%%%%%%%%%%%%%%%%%%%%%%%%%%%%%%%%%%%%%%%%%%%%%%%%%%%%%%%%%%%%%%%%%%%%%%%%%%%%%%%%%%%%%%%%%%%%%%%%%%%%%%%%%%%%%%%%%%%%%%%%%%%%%%%%%%%%%%%%%%%%%
\begin{figure}[h!]
\includegraphics[width=80mm, angle=0,clip]{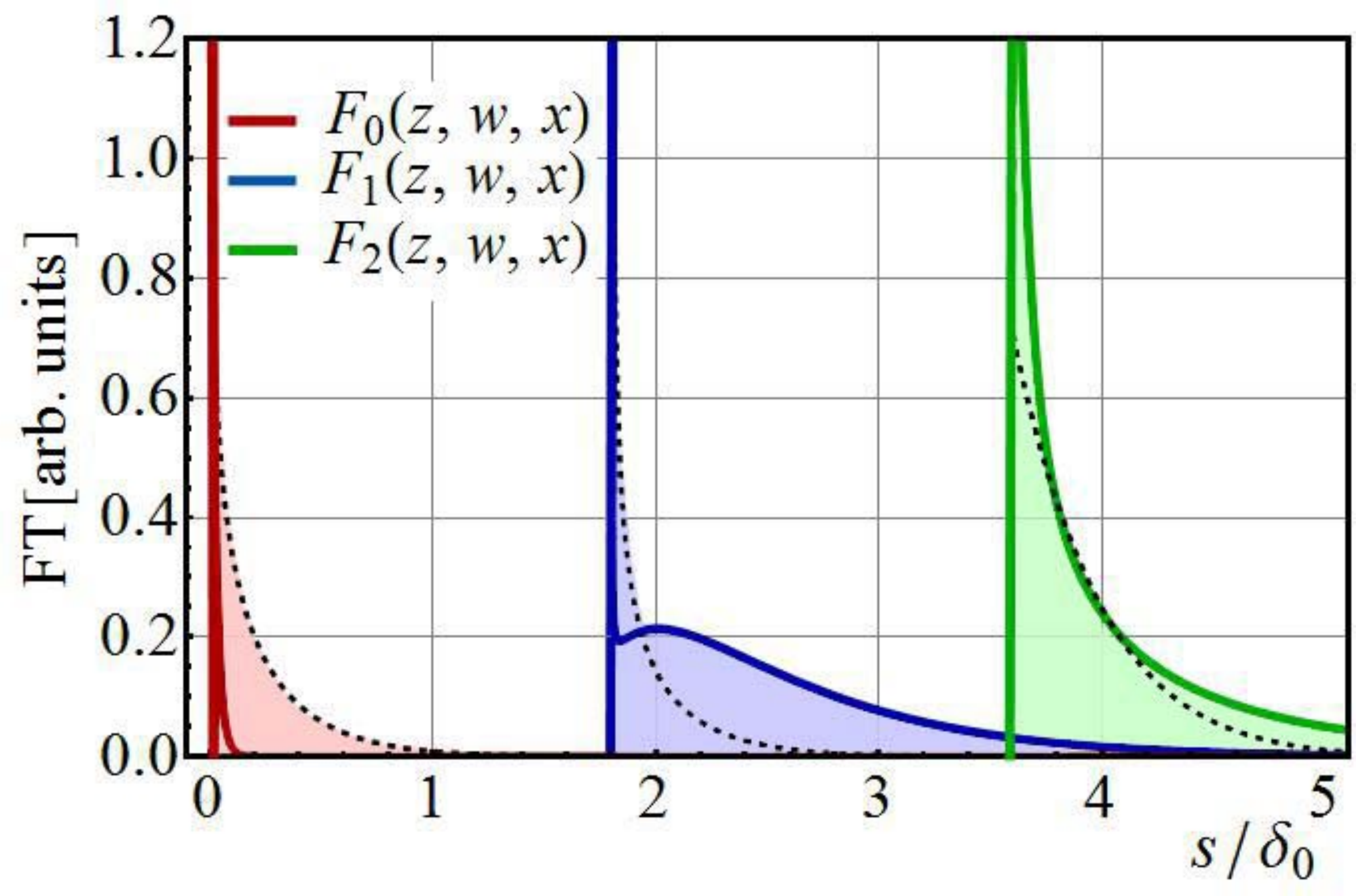}
\vspace{-0.4cm}
\caption{Illustration of the effect of intrapair correlation of disorder on the shapes of the three peaks
$s\ll\Omega_{\s R}$ (red), $s=\Omega_{\s R}$ (blue), and $s=2\Omega_{\s R}$ (green),
in the Fourier transform of $\Delta\sigma(\tau)$. The dashed lines correspond to uncorrelated disorder
($x=0$), while the full lines correspond to strongly correlated disorder ($x=0.95$).
The dimensionless Rabi frequency is $w=\Omega_{\s R}/\delta_{\s 0}=1.2$.}
\label{detuneneg}
\end{figure}
%%%%%%%%%%%%%%%%%%%%%%%%%%%%%%%%%%%%%%%%%%%%%%%%%%%%%%%%%%%%%%%%%%%%%%%%%%%%%%%%%%%%%%%%%%%%%%%%%%%%%%%%%%%%%%%%%%%%%%%%%%%%%%%%%%%%%%%%%%%%%%%%%%%%%%%%%%%%%%%%%%%%%%%%%%%%%%%%%%%%%%%%%%%%%%%

%%%%%%%%%%%%%%%%%%%%%%%%%%%%%%%%%%%%%%%%%%%%%%%%%%%%%%%%%%%%%%%%%%%%%%%%%%%%%%%%%%%%%%%%%%%%%%%%%%%%%%%%%%%%%%%%%%%%%%%%%%%%%%%%%%%%%%%%%%%%%%%%%%%%%%%%%%%%%%%%%%%%%%%%%%%%%%%%%%%%%%%%%%%%%%%
%\begin{figure} [b]
%\includegraphics[width=78mm, angle=0,clip]{Sin_cos.jpg}
%\vspace{-0.4cm}
%\caption{Solid lines: the Fourier transform of the central peak
%plotted using the definition $\bigl[ {\bf\mathrm{F}}^2(s) +
%\tilde{\bf\mathrm{F}}^2(s)\bigr]^{1/2}$ from Eqs.  \eqref{Fourier}, \eqref{Fourier1}
%for two values of the dimensionless Rabi frequency $w=\Omega_{\s R}/\delta_{\s 0}$. Dashed lines %show the corresponding Fourier
%transform plotted using $\bf\mathrm{F}(s)$ only.}
%\label{Fig:sinecosine}
%\end{figure}
%%%%%%%%%%%%%%%%%%%%%%%%%%%%%%%%%%%%%%%%%%%%%%%%%%%%%%%%%%%%%%%%%%%%%%%%%%%%%%%%%%%%%%%%%%%%%%%%%%%%%%%%%%%%%%%%%%%%%%%%%%%%%%%%%%%%%%%%%%%%%%%%%%%%%%%%%%%%%%%%%%%%%%%%%%%%%%%%%%%%%%%%%%%%%%%
Fig. \ref{detuneneg} illustrates how the intrapair correlation of disorder affects the shapes of the peaks in the strong-driving regime. One can see that the prime effect of correlation is a dramatic narrowing of the low--frequency peak. This narrowing reflects the fact that the  low--frequency peak is entirely due to inequivalences of the pair partners, which are suppressed by the correlation. Another effect of strong correlation is that ${\bf \mathrm{F}}_{\s 1}(s)$ develops a maximum at $(s-\Omega_{\s R})={\s \Delta}s/2$ and a minimum at $(s-\Omega_{\s R})=(1-x){\s \Delta}s$. The origin of the maximum is that, for full correlation ($x=1$) the portion of resonant pairs in which only one partner participates to the Rabi oscillations vanishes. More precisely, ${\bf \mathrm{F}}_{\s 1}(s)\propto (s-\Omega_{\s R})^{1/2}\exp{\left[-\frac{s-\Omega_{\s R}} {{\s \Delta} s}\right]}$ at $x=1$, giving rise to a maximum. For small but finite $(1-x)$ such pairs exist and cause ``normal" divergent behavior ${\bf \mathrm{F}}_{\s 1}(s)\propto \frac{1-x}{\sqrt{s-\Omega_{\s R}}}$ at $s\rightarrow \Omega_{\s R}$, resulting in a minimum. Finally, the $s=2\Omega_{\s R}$ peak becomes enhanced by correlation near the origin for the same reason why ${\bf \mathrm{F}}_{\s 1}(s)$ gets
depleted. Conservation of the total area, which applies for all three peaks,
is achieved due to depletion in the body.

\begin{figure}
\includegraphics[width=78mm, angle=0,clip]{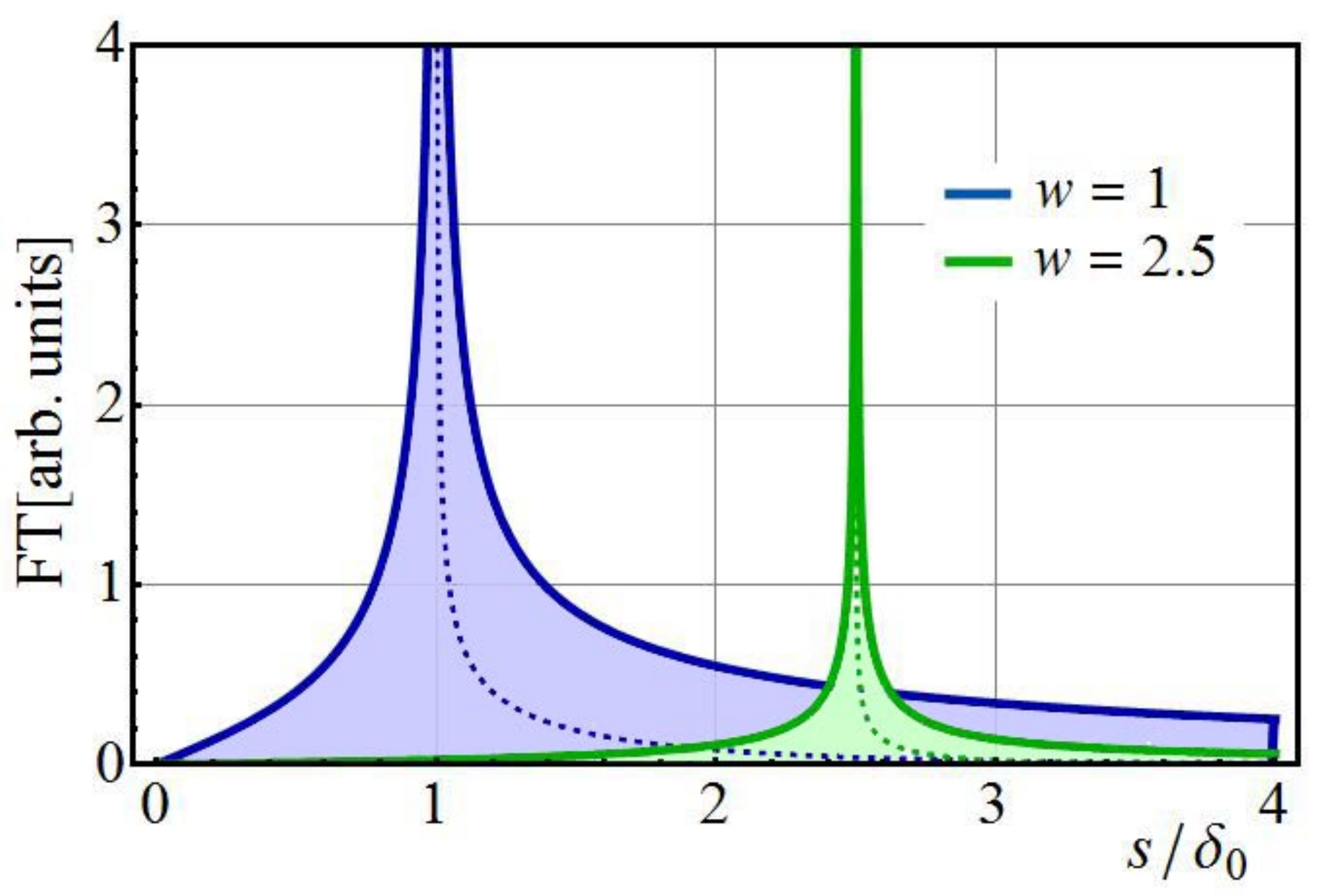}
\vspace{-0.4cm}
\caption{Solid lines: the Fourier transform of the central peak
plotted using the definition $\bigl[ {\bf\mathrm{F}}^2(s) +
\tilde{\bf\mathrm{F}}^2(s)\bigr]^{1/2}$ from Eqs.  \eqref{Fourier}, \eqref{Fourier1}
for two values of the dimensionless Rabi frequency $w=\Omega_{\s R}/\delta_{\s 0}$. The dashed lines show the corresponding Fourier
transform plotted using $\bf\mathrm{F}(s)$ only.}
\label{Fig:sinecosine}
\end{figure}

We note that disorder broadens the peaks in $\bf\mathrm{F}(s)$ only to the right (the higher frequencies) from corresponding thresholds. This is due to the adopted definition [Eq. (\ref{Fourier})] of the Fourier transform. If we used the standard definition, $\bigl[ {\bf\mathrm{F}}^2(s) + \tilde{\bf\mathrm{F}}^2(s)\bigr]^{1/2}$, where $\tilde{\bf\mathrm{F}}(s)$ is defined as
\begin{align}
\label{Fourier1}
&{\tilde{\bf\mathrm{F}}}(s)=\int_0^\infty d\tau\sin(s\tau)
\Big\langle \big(\Delta\sigma(\tau)-\Delta\sigma(0)\big)\Big\rangle, &
\end{align}
the disorder will broaden the peaks both to the left and to the right (to higher and lower frequencies) from the threshold. This is illustrated in Fig.~\ref{Fig:sinecosine} for the central peak at $s=\Omega_{\s R}$. A formal expression
\begin{flalign}
\label{tilde F1 nocorr}
&{\tilde{\bf \mathrm{F}}}_{\s 1}(s)=\left(\frac{\Omega_{\s R}^3}{2\pi\delta_{\s 0}^2s}\right)
\int d\delta~\frac{e^{-\delta^2/2\delta_{\s 0}^2}}{s^2-\Omega_{\s R}^2-\delta^2}~
f\left(\frac{\Omega_{\s R}^2}{2\delta_{\s 0}^2}\right)\!&
\end{flalign}
emerges as an obvious modification of Eq. (\ref{F1 nocorr}). This integral in Eq. (\ref{tilde F1 nocorr}) can be expressed through the error-function. It is nonzero for $s<\Omega_{\s R}$ and for $s>\Omega_{\s R}$. Near $s=\Omega_{\s R}$ it diverges as $\vert s-\Omega_{\s R}\vert^{-1/2}$. As shown in Fig.~\ref{Fig:sinecosine}, the inclusion of $\tilde{\bf\mathrm{F}}(s)$ into the Fourier transform (as done for most previously published pEDMR and pODMR Rabi spectroscopy studies) leads to a significant  unnecessary increase of the peak width which complicates the analysis of experimental data.
%%%%%%%%%%%%%%%%%%%%%%%%%%%%%%%%%%%%%%%%%%%%%%%%%%%%%%%%%%%%%%%%%%%%%%%%%%%%%%%%%%%%%%%%%%%%%%%%%%%%%%%%%%%%%%%%%%%%%%%%%%%%%%%%%%%%%%%%%%%%%%%%%%%%%%%%%%%%%%%%%%%%%%%%%%%%%%%%%%%%%%%%%%%%%%%
%%%%%%%%%%%%%%%%%%%%%%%%%%%%%%%%%%%%%%%%%%%%%%%%%%%%%%%%%%%%%%%%%%%%%%%%%%%%%%%%%%%%%%%%%%%%%%%%%%%%%%%%%%%%%%%%%%%%%%%%%%%%%%%%%%%%%%%%%%%%%%%%%%%%%%%%%%%%%%%%%%%%%%%%%%%%%%%%%%%%%%%%%%%%%%%

\begin{figure}
\includegraphics[width=78mm, angle=0,clip]{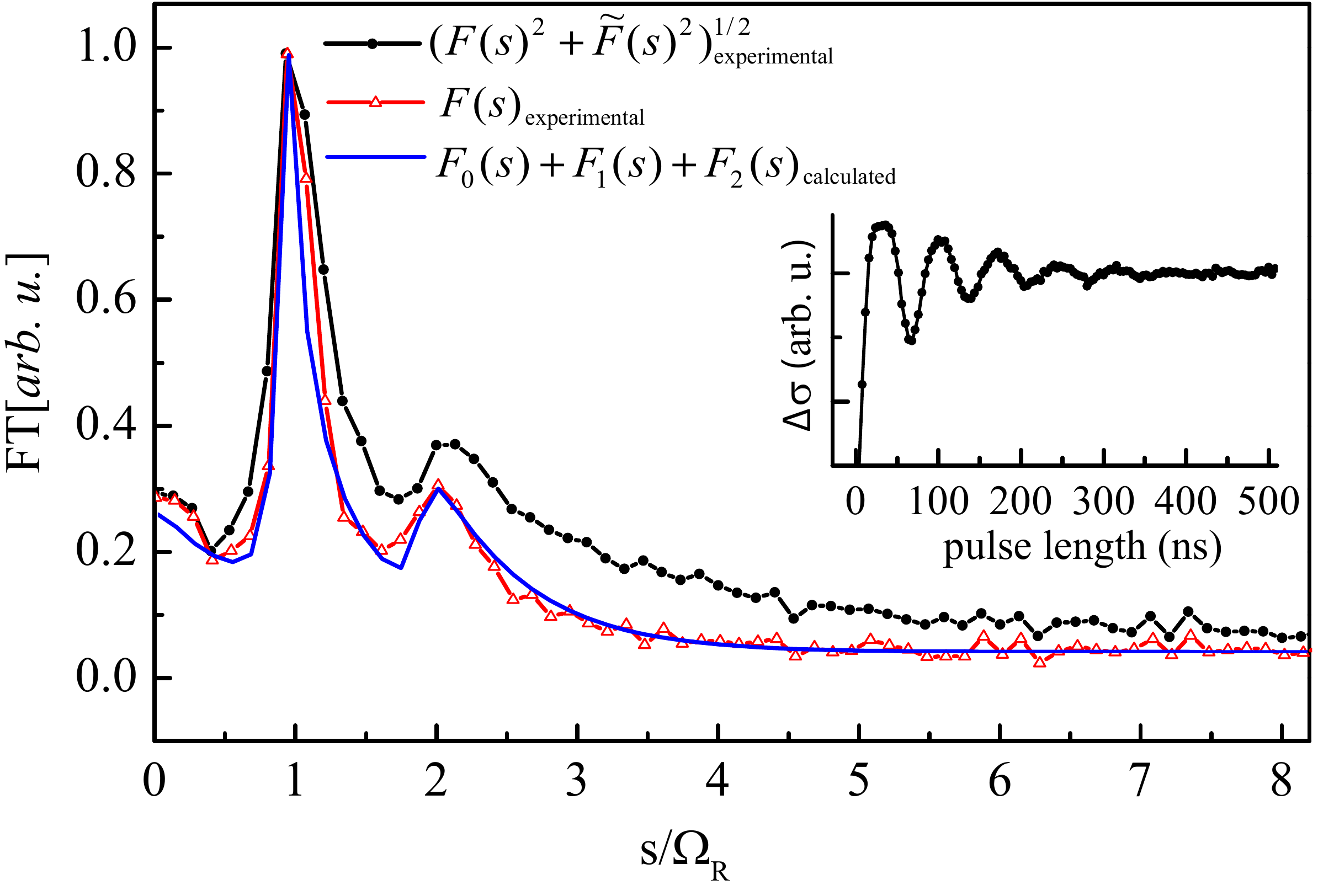}
\vspace{-0.4cm}
\caption{Fourier transformation of an electrically detected spin--Rabi oscillation measured in an organic polymer diode. The inset displays the measured data set, for details about the measurement see text as well as Ref.~\onlinecite{PhysRevLett.108.267601}. The black data of the main panel displays is the {\em absolute} Fourier transform $\sqrt{F(s)^2+\tilde{F}(s)^2}$ of the data in the inset. The red data displays is the {\em real part} $F(s)$ of the Fourier transform of the data in the inset. The blue line represents a Fit of the experimental data with the analytical function $F(s)=F_0(s)+F_1(s)+F_2(s)$}.
\label{Fig:exdata}
\end{figure}

For the illustration of the artificial broadening effect, we refer to an experimental data set which is displayed in the inset of Fig.~\ref{Fig:exdata}. The data shows electrically detected spin--Rabi oscillation in a $\pi$-conjugated polymer diode measured at room temperatures on identically prepared samples and conditions as for the experiments described in Ref.~\onlinecite{PhysRevLett.108.267601}. The diode consisted of an ITO/PEDOT/MEH-PPV/Ca/Al device stack. The signal was measured by detection of changes to a forward steady state current of $I=100\mu$A current, under application of a 4V bias at room temperature. The experiment was conducted under an applied magnetic field of 344mT, the applied magnetic resonant excitation had a frequency of 9.6606GHz and a driving field strength $B_1=0.519(36)$mT produced by a 27W coherent pulsed microwave source. The inset of Fig.~\ref{Fig:exdata} shows the measured raw data. It displays a rapidly dephasing oscillation which is due to inhomogeneities of the applied $B_1$ field, Ref. ~\onlinecite{PhysRevLett.108.267601}. The main plot of Fig.~\ref{Fig:exdata} shows two different Fourier transformations of the data displayed in the inset plotted on a normalized scale as a function of frequency in units of $\Omega_{\s R}=\gamma B_{\s 1}$. The black plot displays the absolute Fourier transform of the experimental data while the red plot displays one the in-phase ($\cos$) component of the Fourier transform. From these plots, it becomes evident that the absolute Fourier transform displays significant broadening of the Rabi spectrum without providing additional insight. It shall be noted that the peak widths of the plot in Fig.~\ref{Fig:exdata} are too close to the frequency resolution of the Fourier transform to unambiguously identify that the peaks of the red plot are more asymmetric compared to the black plot. With the given length of the original data set (500ns), an additional symmetric broadening effect is introduced to both Fourier transforms which overlaps the intrinsic peak shapes. However, for the data sets displayed in Fig.~\ref{Fig:exdata}, one can conclude, that the analysis of experimental spin--Rabi spectra measured with pEDMR and pODMR should be conducted on Fourier data obtained from Eq.~\eqref{Fourier}.

%%%%%%%%%%%%%%%%%%%%%%%%%%%%%%%%%%%%%%%%%%%%%%%%%%%%%%%%%%%%%%%%%%%%%%%%%%%%%%%%%%%%%%%%%%%%%%%%%%%%%%%%%%%%%%%%%%%%%%%%%%%%%%%%%%%%%%%%%%%%%%%%%%%%%%%%%%%%%%%%%%%%%%%%%%%%%%%%%%%%%%%%%%%%%%%
%%%%%%%%%%%%%%%%%%%%%%%%%%%%%%%%%%%%%%%%%%%%%%%%%%%%%%%%%%%%%%%%%%%%%%%%%%%%%%%%%%%%%%%%%%%%%%%%%%%%%%%%%%%%%%%%%%%%%%%%%%%%%%%%%%%%%%%%%%%%%%%%%%%%%%%%%%%%%%%%%%%%%%%%%%%%%%%%%%%%%%%%%%%%%%%

In order to scrutinize the  analytical description of the Rabi-spectrum of a pair process given above, we can now fit the experimentally obtained Fourier transform of the electrically detected electron spin--pair Rabi--oscillation displayed in Fig.~\ref{Fig:exdata} with the analytical expressions in Eqs.~\eqref{F1 nocorr}, \eqref{SecondPeak}, and \eqref{smallS} for the Fourier transforms of an uncorrelated spin pair ensemble $F_1(s)$, $F_2(s)$, and $F_0(s)$, respectively. It is known that weakly coupled polaron pairs in MEH-PPV do not exhibit a correlation of their local hyperfine fields~\cite{BoehmeMain} so an agreement between the experimental data set and the sum of Eqs.~\eqref{F1 nocorr}, \eqref{SecondPeak}, and \eqref{smallS} is expected. Since the expressions for $F_i(s)$ are absolute, the fit of $F(s)=F_0(s)+F_1(s)+F_2(s)$ has only three fit parameters, namely (i) a global scaling factor of the unitless ordinate, (ii) the disorder factor $\Omega_{\s R}/\delta_{\s 0}$ as well as a third fit variable which is the width of a homogenous (Lorentzian) broadening of $F(s)$. This broadening is due to an experimental artifact, coming from a Fourier transformation of a finite time interval as well as the measured decay of the Rabi--oscillation which is known to be due to inhomogeneities of the experimentally generated $B_1$ field~\cite{PhysRevLett.108.267601}. Thus, the length (500ns) of the experimentally recorded Rabi oscillation determine the frequency resolution of the resulting Fourier transform while the decay ($\approx$ 150ns) of the Rabi--oscillation further contributed to homogeneous broadening of the experimental Fourier transformed data.

Using the three fit variables for $F(s)$ mentioned above, we obtain a good fit of the experimental data as displayed by the blue line in Fig.~\ref{Fig:exdata}. In this fit, the disorder parameter $\delta_{\s 0}/\Omega_{\s R}=1.54$ with an estimated error of less than 0.2. Given the Rabi nutation frequency $\Omega_{\s R}=\gamma B_{\s 1}=14.5(1.0) $MHz, we obtain $\delta_{\s 0}=22.3(3.3)$MHz. This value is equivalent to a hyperfine broadened polaron  line width of $\delta_{\s 0}/\gamma=0.80(14)$mT, a value that is in good agreement with the widths of EDMR detected polaron spin resonances in MEH-PPV~\cite{BoehmeMain,wrong,PhysRevLett.108.267601}. We conclude from this that the analytical expressions  Eqs.~\eqref{F1 nocorr}, \eqref{SecondPeak}, and \eqref{smallS} capture adequately the Rabi oscillations in weakly coupled spin pairs and should therefore be used for the fit of pEDMR data governed by non--correlated spin pair ensembles.
\\

%%%%%%%%%%%%%%%%%%%%%%%%%%%%%%%%%%%%%%%%%%%%%%%%%%%%%%%%%%%%%%%%%%%%%%%%%%%%%%%%%%%%%%%%%%%%%%%%%%%%%%%%%%%%%%%%%%%%%%%%%%%%%%%%%%%%%%%%%%%%%%%%%%%%%%%%%%%%%%%%%%%%%%%%%%%%%%%%%%%%%%%%%%%%%%% %%%%%%%%%%%%%%%%%%%%%%%%%%%%%%%%%%%%%%%%%%%%%%%%%%%%%%%%%%%%%%%%%%%%%%%%%%%%%%%%%%%%%%%%%%%%%%%%%%%%%%%%%%%%%%%%%%%%%%%%%%%%%%%%%%%%%%%%%%%%%%%%%%%%%%%%%%%%%%%%%%%%%%%%%%%%%%%%%%%%%%%%%%%%%%% %%%%%%%%%%%%%%%%%%%%%%%%%%%%%%%%%%%%%%%%%%%%%%%%%%%%%%%%%%%%%%%%%%%%%%%%%%%%%%%%%%%%%%%%%%%%%%%%%%%%%%%%%%%%%%%%%%%%%%%%%%%%%%%%%%%%%%%%%%%%%%%%%%%%%%%%%%%%%%%%%%%%%%%%%%%%%%%%%%%%%%%%%%%%%%% %%%%%%%%%%%%%%%%%%%%%%%%%%%%%%%%%%%%%%%%%%%%%%%%%%%%%%%%%%%%%%%%%%%%%%%%%%%%%%%%%%%%%%%%%%%%%%%%%%%%%%%%%%%%%%%%%%%%%%%%%%%%%%%%%%%%%%%%%%%%%%%%%%%%%%%%%%%%%%%%%%%%%%%%%%%%%%%%%%%%%%%%%%%%%%%

\begin{acknowledgements}
We acknowledge D. P. Waters and R. Baarda for the preparation of the MEH-PPV diode pEDMR templates.
%and W. J. Baker for the fabrication of the MEH-PPV diodes as well as for conducting the pEDMR %measurement and the processing of the experimental data.
We also acknowledge the support of this work by the National Science Foundation through the Materials Research Science and Excellence Center (\#DMR-1121252). CB further acknowledges the support through a National Science Foundation CAREER award (\#0953225).
\end{acknowledgements}


\begin{thebibliography}{3}

\bibitem{PhysRevB.79.195205}
 T. W. Herring, S.-Y. Lee, D. R. McCamey, P. C. Taylor,
K. Lips, J. Hu, F. Zhu, A. Madan, and C. Boehme, Phys. Rev. B {\bf 79}, 195205 (2009).
\bibitem{lee_192104}
S.-Y. Lee, S.-Y. Paik,  D. R. McCamey, J. Hu,  F.  Zhu,
A. Madan, and C. Boehme, Phys. Rev. Lett.  {\bf 97}, 192104
 (2010).

\bibitem{Boehme2003b} C. Boehme and K.  Lips,  Phys. Rev. Lett.  {\bf 91},  246603
(2003).
\bibitem{Stegner2006}
A.  R. Stegner, Nature Phys. {\bf 2}, 835 (2006).

\bibitem{PhysRevLett.106.187601}
F. Hoehne, L. Dreher, H. Huebl, M. Stutzmann, and M. S.
Brandt,  Phys. Rev. Lett.  {\bf 106}, 187601 (2011).

\bibitem{PhysRevLett.100.177602}
H. Huebl, F. Hoehne, B. Grolik,  A. R. Stegner, M. Stutz-
mann, and M. S. Brandt,  Phys. Rev. Lett.  {\bf 100},  177602 (2008).

\bibitem{BoehmeC60}
W. Harneit, C. Boehme, S. Schaefer, K. Huebener, K. Fo-
stiropoulos,  and K.  Lips,  Phys. Rev. Lett.  {\bf 98},  216601 (2007).

\bibitem{BoehmeNature}
D. R. McCamey, H. A. Seipel, S.-Y. Paik, M. J. Walter,
N. J. Borys, J. M. Lupton, and C. Boehme, Nat. Mater. {\bf 7},
723 (2008).
 %%%%%%%%%%%%%%%%%%%%%%%%%%%%%%%%%%%%%%%%
\bibitem{schaefer2008electrical}
S. Schaefer, S. Saremi, K. Fostiropoulos,  J.  Behrends,
K.  Lips,  and W.  Harneit, Phys. Stat. Sol. (b)  {\bf 245}, 2120 (2008).

\bibitem{BoehmeDifferentiation}
W. J. Baker, D. R. McCamey, K. J. van Schooten, J. M.
Lupton, and C. Boehme, Phys. Rev. B {\bf 84}, 165205 (2011).

\bibitem{wrong}
J. Behrends, A. Schnegg, K. Lips, E. A. Thomsen, A. K.
Pandey, I. D. W. Samuel, and D. J. Keeble, Phys. Rev.
Lett.  {\bf 105}, 176601 (2010).

\bibitem{BoehmeMain}
D. R. McCamey, K. J. van Schooten, W. J. Baker, S.-Y. Lee, S.-Y. Paik, J. M. Lupton, and C. Boehme, Phys. Rev. Lett. {\bf 104}, 017601 (2010).

\bibitem{CPHC_CPHC201000186}
J.   M.   Lupton,    D.   R.   McCamey,   and  C.   Boehme,
Chem. Phys. Chem. {\bf 11}, 3040 (2010).

\bibitem{Mott}
 D. Kaplan, I. Solomon, and N. F. Mott, J. Phys. (Paris) {\bf 39}, L51 (1978).
\bibitem{Boehme1}V. Rajevac, C. Boehme, C. Michel, A. Gliesche, K. Lips, S. D. Baranovskii, and P. Thomas, Phys. Rev. B {\bf 74}, 245206 (2006).

\bibitem{Boehme2}A. Gliesche, C. Michel, V. Rajevac, K. Lips, S. D. Baranovskii, F. Gebhard, and C. Boehme, Phys. Rev. B {\bf 77}, 245206 (2008).
%Exchange

\bibitem{Boehme3}C. Michel, A. Gliesche, S. D. Baranovskii, K. Lips, F. Gebhard, and C. Boehme, Phys. Rev. B {\bf 79}, 052201 (2009).
%Role of disorder and correlation
\bibitem{Boehme0}
C. Boehme and K. Lips, Phys. Rev. B {\bf 68}, 245105 (2003).

\bibitem{Loss}
   F. Koppens, D. Klauser, W. Coish, K. Nowack, L. Kouwenhoven, D. Loss, and L. Vandersypen, Phys. Rev. Lett. {\bf 99}, 106803 (2007).

\bibitem{spaeth2003point}
J. Spaeth and H. Overhof, {\em Point  defects in  semiconduc-
tors and insulators: determination of atomic and electronic structure from paramagnetic hyperfine
interactions}, {\bf 51} (Springer Verlag, 2003).

\bibitem{nguyen2010isotope}
T.  Nguyen,  G.  Hukic-Markosian,  F.  Wang,  L.  Wojcik,
X. Li, E. Ehrenfreund, and Z. Vardeny, Nat. Mater.
{\bf 9}, 345 (2010).

\bibitem{PhysRevLett.106.037601}
L. Dreher, T. A. Hilker, A. Brandlmaier, S. T. B. Goennenwein, H. Huebl,
M. Stutzmann, and M. S. Brandt,  Phys. Rev. Lett. {\bf 106}, 037601 (2011).


\bibitem{PhysRevLett.108.267601}
W. J. Baker, T. L. Keevers, J. M. Lupton, D. R. McCamey,
and C. Boehme, Phys. Rev. Lett.  {\bf 108}, 267601 (2012).



\end{thebibliography}
\end{document}